\documentclass{svjour3}
\journalname{Journal of Statistical Physics}
\usepackage{mathptmx}
\usepackage{amsmath}
\usepackage{amssymb}
\usepackage{multirow}
\usepackage{graphicx}
\def\tw{t_\mathrm{w}}
\newcommand{\vn}[1]{\textit{\bfseries #1}}

\begin{document}

\title{An in-depth view of the microscopic dynamics of Ising spin glasses at
fixed temperature}

\author{F.~Belletti \and A.~Cruz \and L.A.~Fernandez \and A.~Gordillo-Guerrero
\and M.~Guidetti \and A.~Maiorano \and F.~Mantovani
\and E.~Marinari \and V.~Martin-Mayor \and J. Monforte \and A.~Mu\~noz~Sudupe
\and D.~Navarro \and G.~Parisi \and S.~Perez-Gaviro
\and J.J.~Ruiz-Lorenzo \and S.F.~Schifano \and D.~Sciretti
\and A.~Tarancon \and R.~Tripiccione \and D.~Yllanes}

\institute{
F.~Belletti \and M.~Guidetti \and A.~Maiorano \and F.~Mantovani \and
S.F.~Schifano \and R.~Tripiccione \at
Dipartimento di Fisica Universit\`a di Ferrara and
INFN - Sezione di Ferrara, Ferrara, Italy.
\and
A.~Cruz \and L.A.~Fernandez \and A.~Gordillo-Guerrero
\and A.~Maiorano \and V.~Martin-Mayor \and J. Monforte \and  S.~Perez-Gaviro \and
J.J.~Ruiz-Lorenzo \and D.~Sciretti \and A.~Tarancon
\and D.~Yllanes\at
Instituto de Biocomputaci\'on y F\'{\i}sica de
Sistemas Complejos (BIFI), Zaragoza, Spain
\and
A.~Cruz \and  A.~Tarancon \at
Departamento de F\'\i{}sica Te\'orica, Universidad
  de Zaragoza, 50009 Zaragoza, Spain.
\and
L.A.~Fernandez \and V.~Martin-Mayor \and A.~Mu\~noz~Sudupe
\and D.~Yllanes\at
Departamento de F\'\i{}sica Te\'orica I, Universidad
  Complutense, 28040 Madrid, Spain.
\and
A.~Gordillo-Guerrero  \at
Dpto. de Ingener\'{\i}a El\'ectrica, Electr\'onica y Autom\'atica,
Universidad de Extremadura.
 Avda. de la Universidad s/n. 10071. C\'aceres, Spain.
\and
A.~Maiorano \and E.~Marinari \and G.~Parisi \and S.~Perez-Gaviro\at
Dipartimento di Fisica, SMC of INFM-CNR and
  INFN, Universit\`a di Roma {\it La Sapienza}, 00185 Roma, Italy.
\and
D.~Navarro\at
Departamento de Ingenier\'{\i}a, Electr\'onica y Comunicaciones
and Instituto de Investigaci{\'o}n en Ingenier{\'i}a de Arag{\'o}n (I3A),
Universidad de Zaragoza, 50018 Zaragoza, Spain.
\and
J.J.~Ruiz-Lorenzo \at
Departamento de
  F\'{\i}sica, Universidad de Extremadura, 06071 Badajoz, Spain.
}
\maketitle
\authorrunning{Janus Collaboration}

\begin{abstract}
  Using the dedicated computer Janus, we follow the nonequilibrium dynamics of the
  Ising spin glass in three dimensions for eleven orders of magnitude.
  The use of integral estimators for the coherence and correlation lengths allows
  us to study dynamic heterogeneities and the presence of a replicon
  mode and to obtain safe bounds on the Edwards-Anderson order
  parameter below the critical temperature.  We
  obtain good agreement with
  experimental determinations of the temperature-dependent decay
  exponents for the thermoremanent magnetization.
  This magnitude is observed to scale with the much harder
  to measure coherence length, a potentially useful
  result for experimentalists. The exponents for energy
  relaxation display a  linear dependence on
  temperature and reasonable extrapolations to the critical point. 
  We conclude examining the time growth of the
  coherence length, with a comparison of critical and activated dynamics.

\end{abstract}

\section{Introduction}
Below their glass temperature, Spin Glasses~\cite{EXPBOOK} (SG) are
perennially out of equilibrium. The understanding of their
sophisticated dynamical behavior is a long standing challenge both to
theoretical and to experimental physics.

Aging~\cite{LETDYN1} is a feature of SG dynamics that shows up even in
the simplest experimental protocol, the {\em direct quench}.  In these
experiments, the SG is cooled as fast as possible to the working
temperature below the critical one, $T<T_\mathrm{c}$. It is then let to
equilibrate for a {\em waiting time}, $t_\mathrm{w}$, its properties
to be probed at a later time, $t+t_\mathrm{w}$. For instance one may
cool the SG in the presence of an external field, which is switched
off at time $\tw$. The so-called thermoremanent magnetization decays
with time, but the larger $\tw$ is, the slower the decay. In fact,
it has been claimed that, if the cooling is fast enough, the
thermoremanent magnetization depends upon $t$ and $t_\mathrm{w}$
only through the combination $t/t_\mathrm{w}$, at least for
$10^{-3}<t/t_\mathrm{w}<10 $ and $t_\mathrm{w}$ in the range
50\,s\,---\,$10^4$\,s~\cite{RODRIGUEZ}. In other words, the only
characteristic time scale is the sample's own age as a glass, $\tw$
(this behavior is named Full Aging).  Note, however, that there is
some controversy regarding the natural time variable which could
rather be $t/t_\mathrm{w}^\mu$ with $\mu$ slightly less than
one~\cite{SACLAY}.

The time evolution is believed to be caused by the growth of coherent
spatial domains. Great importance is ascribed to the size of these
domains, the coherence length $\xi(\tw)$, which is accessible to
experiments through estimates of Zeeman energies~\cite{ORBACH}. The
time evolution of $\xi(\tw)$ plays a crucial role in the droplets
theory of SG nonequilibrium isothermal
dynamics~\cite{SUPERUNIVERSALITY}. Perhaps unsurprisingly, it also
plays a central role in yet incipient attempts to rationalize memory
and rejuvenation experiments (see~\cite{MR-EXPERIMENTS,BERT,BERBOU,SUE3} and references therein), where
the experimentalist probes the glassy state by playing with the
working temperature.

Even for the simplest direct quench experiment, there is some polemics
regarding the growth law of $\xi(\tw)$: some theories advocate a
logarithmic growth~\cite{SUPERUNIVERSALITY}, while a power law describes
 numerical simulations~\cite{MPRTRL_1999} or
experiments~\cite{ORBACH} better (a somewhat intermediate scaling has been
proposed by the Saclay group~\cite{BOUCHAUD-SACLAY} and found useful in
experimental work~\cite{BERT}, see also Sect.~\ref{XIGROWTH} below).
Nevertheless, two facts are firmly established: (i) the lower $T$ is,
the slower the growth of $\xi(t_\mathrm{w})$ and (ii) $\xi\sim 100$
lattice spacings, even for $T\sim T_\mathrm{c}$ and $t_\mathrm{w}$ as
large as $10^4$~s~\cite{ORBACH}.  Hence, the study of SG in thermal
equilibrium seems confined to nanometric samples, or to numerical
simulations.

There is clear evidence, both experimental~\cite{EXPERIMENTOTC} and
theoretical~\cite{BALLESTEROS,PALASS-CARACC}, for a thermodynamic
origin of this sluggish dynamics. A SG phase appears below the critical
temperature, $T_\mathrm{c}$. Several theories propose mutually
contradicting scenarios for the {\em equilibrium} SG phase: the
droplets~\cite{DROPLET}, replica symmetry breaking (RSB)~\cite{RSB}, and
the intermediate Trivial-Non-Trivial (TNT) picture~\cite{TNT}. Even if
this equilibrium phase is experimentally unreachable (at least in
human time scales), we now know~\cite{FRANZ} that it is nevertheless relevant to
the nonequilibrium dynamics probed by experiments.

Droplets expects two equilibrium states related by global spin
reversal. The SG order parameter, the spin overlap $q$ (precise
definitions are given below in Sect.~\ref{DEF-SECT}), takes only two values
$q=\pm q_\mathrm{EA}\,.$ In the RSB scenario an infinite number of
pure states influence the dynamics~\cite{RSB,FEGI,CONTUCCI-SEP}, so
that all $-q_\mathrm{EA}\!\leq\!q\!\leq\!q_\mathrm{EA}$ are reachable.
TNT~\cite{TNT} describes the SG phase similarly to an antiferromagnet
with random boundary conditions: even if $q$ behaves as for RSB
systems, TNT agrees with droplets in the vanishing surface to volume
ratio of the largest thermally activated spin domains (i.e. the
link-overlap defined below takes a single value).

Droplets' isothermal aging~\cite{SUPERUNIVERSALITY} is that of a
disguised ferromagnet.\footnote{However, when the temperature is
  varied, droplets theory predicts a more complex behavior for SGs
  than for ferromagnets, due to temperature chaos.} Indeed,
superuniversality, the emerging picture of isothermal aging, has been
found useful for the study of basically all coarsening systems.  For
$T<T_\mathrm{c}\,$ the growing domains are compact geometrical
objects. Even if the surface of these domains might be fractal, their
surface to volume ratio vanishes as $\xi(t_\mathrm{w})$ diverges, see
Eq.~(\ref{C4-DECAY}) below. Inside them, the spin overlap
coherently takes one of its possible equilibrium values $q=\pm q_\mathrm{EA}$.
Time dependencies are entirely encoded in the growth law of these
domains, since correlation functions (in principle depending on time
and distance, $r$) are universal functions of $r/\xi(t_\mathrm{w})$.

We are not aware of any investigation of the dynamical consequences of
the TNT picture. Nevertheless, the antiferromagnet analogy suggests that
TNT systems will show coarsening behavior.

As for the RSB scenario, equilibrium states with a vanishing order
parameter $q\!=\!0$ do exist. Hence, the nonequilibrium dynamics
starts, and remains forever, with a vanishing order parameter.
Furthermore, the replicon, a Goldstone mode analogous to magnons in
Heisenberg ferromagnets, is present for all
$T<T_\mathrm{c}$~\cite{DeDominicis}. As a consequence, the spin overlap
is expected to vanish {\em inside  each domain} in the limit of a large
$\xi(\tw)$. Furthermore, $q$ is not a privileged observable (overlap
equivalence~\cite{FEGI}): the link overlap displays equivalent Aging
behavior.

In order to be quantitative, these theoretical pictures of
nonequilibrium dynamics need numerical computations for model
systems. Indeed, several investigations have been carried out even for
the simplest cooling protocol, the direct
quench~\cite{BERBOU,SUE3,MPRTRL_1999,KISKER,RIEGER,SUE,SUE2,LET-NUM1,LET-NUM2,LET-NSU}.
A major drawback of this approach, however, is the shortness of the
reachable times. Indeed, one Monte Carlo Step (MCS) corresponds to
$10^{-12}$ s~\cite{EXPBOOK}. The experimental scale is at $10^{14}$ MCS
($\sim 100$\,s), while typical nonequilibrium simulations reach $\sim
10^{-5}$s. The problem has been challenging enough to compel
physicists to design high-performance computers for SG
simulations~\cite{OGIELSKI,SUE_DEF,JANUS}.

The situation has dramatically changed thanks to {\em
  Janus}~\cite{JANUS}, an FPGA computer that allows us to simulate the
dynamics of a reasonably large SG system for eleven orders of
magnitude, from picoseconds to tenths of a second.\footnote{The
  wall-clock time needed for this computation on a cubic lattice of
  $80$ lattice spacings is some 25 days.} Thanks to Janus, we have
recently performed a study of the nonequilibrium dynamics of the Ising
Spin Glass~\cite{DYNJANUS}.  We introduced novel analysis techniques
that allow the computation of the coherence length in a model independent way.
This was crucial to obtain evidence for a replicon
correlator. Furthermore, we showed how to investigate overlap
equivalence and presented evidence for it.

In this work, we shall concentrate on the simplest protocol, the direct
quench, for an Ising SG.  We present a detailed study of dynamic
heterogeneities, an aspect untouched upon in~\cite{DYNJANUS} in spite
of its relevance~\cite{LET-NUM1,LET-NUM2,LET-NSU}. We show the first conclusive
numerical evidence for a growing {\em correlation} length in the
nonequilibrium dynamics, and its relationship with the {\em coherence}
length $\xi(t_\mathrm{w})$ is explored. Furthermore, we compute the
anomalous dimension for the two-time, two-site propagator (see
definitions below). Due to their central role, a systematic way of
extracting coherence (or correlation) lengths from numerical data is
called for, but it has scarcely been investigated in the past (see
however~\cite{SUE3,MPRTRL_1999,KISKER}). This is why we take here the
occasion to give full details on our {\em integral}
estimators~\cite{DYNJANUS} (see also~\cite{KISKER}).

The layout of the rest of this paper is as follows. In
Sect.~\ref{DEF-SECT} we define the model as well as the correlation
functions and time sectors. We describe
our simulations, which have been extended as compared
with~\cite{DYNJANUS} and discuss the difficult topic
of extracting the best fit parameters from extremely correlated data.
Sect.~\ref{XI-SECT} is devoted to the integral
estimators of the coherence (or correlation) lengths. To the best of our
knowledge, the investigation of this technical (but crucial) issue was
started in the context of lattice field theories~\cite{COOPER}. These
integral estimators were instrumental to develop modern Finite Size Scaling
techniques for {\em equilibrium} critical
phenomena~\cite{FSS-PISA,QUOTIENTS} and, therefore,  to
establish the existence of the Spin Glass phase in three
dimensions~\cite{BALLESTEROS,PALASS-CARACC,HEISENBERG}. In the context
of nonequilibrium dynamics, new aspects (and opportunities) appear.
In Sect.~\ref{XIC2TW-SECT} we investigate the dynamic heterogeneities. In
Sect.~\ref{TIME-CORRELATION-FUNCTIONS-SECT}, the information gathered
on length scales is used to analyze time correlation functions (and to
extrapolate them to infinite time). We also study
the thermoremanent magnetization. The crucial issue of the time
growth of the coherence length $\xi(\tw)$ is considered in
Sect.~\ref{XIGROWTH}.
We present our conclusions in Sect.~\ref{CONCLUSIONES}

\section{Model, correlation functions, time sectors}\label{DEF-SECT}
\subsection{Model}

The $D\!=\!3$ Edwards-Anderson Hamiltonian is defined in terms of two
types of degrees of freedom: dynamical and {\em quenched}. The
dynamical ones are the Ising spins $\sigma_{\vn{x}}\!=\!\pm1$, which
are placed on the nodes, $\vn{x}$, of a cubic lattice of
linear size $L$ and periodic boundary conditions. The nondynamical (or
quenched) ones are coupling constants assigned to the lattice links
that join pairs of lattice nearest neighbors,
$J_{\vn{x},\vn{y}}$. In this work
$J_{\vn{x},\vn{y}}\!=\!\pm 1$ with $50\%$
probability. Each assignment of the
$\{J_{\vn{x},\vn{y}}\}$ will be named a {\em sample},
and, once it is fixed, it will never be varied~\cite{EXPBOOK}.

The interaction energy for the spins is
\begin{equation}
{\cal H}=-\sum_{\langle \vn{x}, \vn{y}\rangle } J_{\vn{x},\vn{y}} \sigma_{\vn{x}}\, \sigma_{\vn{y}},\label{EA-H}
\end{equation}
($\langle\cdots\rangle$ denotes lattice nearest neighbors).
The choice $J^2=1$ fixes our energy units.  We work in the so-called
quenched approximation: any quantity (either a thermal mean value or
a time dependent magnitude, see below) is supposed to be computed {\em
  first} for a given assignment of the couplings. Only afterwards the
resulting value is averaged over the $J_{\vn{x},\vn{y}}$, which
we denote by $\overline{(\cdots)}$.\footnote{In principle,
one should average many thermal histories for each sample and
then compute the average over the disorder. In practice,
errors can be largely reduced if one generates
two histories per sample but generates more samples, since
both sources of error are statistically independent. The claims on lack of
self-averageness made in Sect.~\ref{SECT-OBSERVABLES} depend
critically on this choice. In fact, were one to simulate an infinite
number of thermal trajectories per sample, the spin glass
susceptibility would be self-averaging.}

The spins evolve in time with  Heat-Bath dynamics (see,
e.g.,~\cite{VICTORAMIT}), which belongs to the universality class of
physical evolution. The starting spin configuration is taken to be
fully disordered, to mimic the experimental {\em direct quench}
protocol.

Note that the Hamiltonian (\ref{EA-H}) has a global $\mathbf{Z}_2$ symmetry (if
all spins are simultaneously reversed
$\sigma_{\vn{x}}\to -\sigma_{\vn{x}}$ the
energy is unchanged). This symmetry gets spontaneously broken in three
dimensions upon lowering the temperature at the SG transition at
$T_\mathrm{c}=1.109(10)$~\cite{PELISSETTO}.\footnote{The alert reader
will recall that the gauge symmetry of Eq.~\eqref{GAUGE-TRANSF}
forbids a spontaneous magnetization. Indeed, see Sect.~\ref{SECT-OBSERVABLES},
one needs to introduce an independently evolving copy of the spin configuration
which promotes the symmetry to $\mathbf{Z}_2\times\mathbf{Z}_2$. This is
the symmetry which is actually spontaneously broken.}

Finally, let us recall that
the average over the coupling constants induces a non dynamical gauge
symmetry~\cite{TOULOUSE77}. Let us choose a random sign per site
$\epsilon_{\vn{x}}=\pm 1\,$. Hence, the energy (\ref{EA-H}) is
invariant under the transformation
\begin{equation}
\begin{array}{rcl}
s_{\vn{x}}&\longrightarrow &\epsilon_{\vn{x}}s_{\vn{x}},\\
J_{{\vn{x}},{\vn{y}}}&\longrightarrow &\epsilon_{\vn{x}}\epsilon_{\vn{y}} J_{{\vn{x}},{\vn{y}}}\label{GAUGE-TRANSF}
\end{array}
\end{equation}
Since the gauge transformed couplings
$\epsilon_{\vn{x}}\epsilon_{\vn{y}} J_{{\vn{x}},{\vn{y}}}$ are
just as probable as the original ones, the quenched mean value of an
arbitrary function of the spins $O(\{s_{\vn x}\})$ is identical to
that of its gauge-average $\sum_{\{\epsilon_{\vn x}=\pm
  1\}}O(\{\epsilon _{\vn x} s_{\vn x}\})/2^{L^D}\,,$ which typically
is an uninteresting constant value. Constructing non trivial gauge-invariant
observables is the subject of the next subsection.

\subsection{Observables}\label{SECT-OBSERVABLES}

A standard way of forming operators that are gauge-invariant under
(\ref{GAUGE-TRANSF}) is to consider real replicas. These are two statistically
independent systems, $\{\sigma_{\vn{x}}^{(1)}\}$ and
$\{\sigma_{\vn{x}}^{(2)}\}$, evolving in time with the very
same set of couplings. Their (obviously gauge-invariant) overlap field
at time $t_\mathrm{w}$ is
\begin{equation}
q_{\vn{x}}(t_\mathrm{w})= \sigma_{\vn{x}}^{(1)}(t_\mathrm{w})
\sigma_{\vn{x}}^{(2)}(t_\mathrm{w})\,.
\end{equation}

A slight modification consists in using just one of the real replicas, say
$\{\sigma_{\vn{x}}^{(1)}\}$, but considering times $\tw$ and
$t+\tw$
\begin{equation}
c_{\vn{x}}(t,t_\mathrm{w})= \sigma_{\vn{x}}^{(1)}(t+t_\mathrm{w})
\sigma_{\vn{x}}^{(1)}(t_\mathrm{w})\,.
\end{equation}
In many of the quantities defined below using
$c_{\vn{x}}(t,t_\mathrm{w})$, one may obviously gain
statistics by averaging over the two real replicas. We have done so
whenever it was possible, but this will not be explicitly indicated.
We consider three types of quantities:
\begin{enumerate}
\item Single-time global quantities:\\
\begin{itemize}
\item Time-dependent energy density ($N=L^D$ is the number of spins in
  the lattice)
\begin{equation}
e(t_\mathrm{w})=-\frac{1}{N}
\sum_{\langle \vn{x}, \vn{y}\rangle } J_{\vn{x},\vn{y}} \sigma_{\vn{x}}(\tw)\, \sigma_{\vn{y}}(\tw)\,.
\end{equation}
Recall that $\langle \vn{x}, \vn{y}\rangle$ indicates
summation restricted to lattice nearest neighbors.

\item The spin glass susceptibility $\chi_\mathrm{SG}(t_\mathrm{w})$ is defined
in terms of the SG order parameter
\begin{equation}
q(t_\mathrm{w})=\frac{1}{N}\sum_{\vn{x}}\ q_{\vn{x}}(\tw) \,.
\end{equation}
Of course, the quenched mean value $\overline{q(t_\mathrm{w})}$
vanishes in the nonequilibrium regime where the system size is much
larger than the coherence length $\xi(t_\mathrm{w})\,$. The susceptibility
\begin{equation}
\chi_\mathrm{SG}(t_\mathrm{w})=N\overline{q^2(t_\mathrm{w})}\,,\label{CHISG-DEF}
\end{equation}
steadily grows with the size of the coherent domains. Note that fluctuation-dissipation relations
imply that $\chi_\mathrm{SG}$ is basically the nonlinear magnetic susceptibility.\\
\end{itemize}

\item Two-times global correlation functions:
\begin{itemize}
\item Spin-spin correlations:
\begin{equation}
C(t,\tw)=\overline{\frac{1}{N}\sum_{\vn{x}}\ c_{\vn x}(t,\tw)}\,.\label{SPIN-SPIN-CORRELATION}
\end{equation}
The function $C(t,\tw)$ carries many meanings:
\begin{enumerate}
\item If the first argument $\tw$ is held fixed, and
  $C(t,t_\mathrm{w})$ is studied as $t$ grows, it is just the
  thermoremanent magnetization. Indeed, because of the symmetry
  (\ref{GAUGE-TRANSF}) the uniform configuration that would have been
  enforced by holding the spin glass in a strong external magnetic
  field can be gauged to the spin configuration found at time
  $t_\mathrm{w}$ after a random start.
\item
On the other hand, in the pseudoequilibrium regime $t\ll t_\mathrm{w}$,
the (real part of the) magnetic susceptibility at frequency  $\omega=2\pi/T$
is given by the fluctuation-dissipation formula $\bigl(1-C(t,t_\mathrm{w})\bigr)/T\,.$
\item
Another point we shall be concerned with is the computation of the SG
order parameter. It may be defined from the translationally
invariant time sector\footnote{By analogy, one may define a
  $\mu$ time sector by the limit $C_\infty^{(\mu)}(s)=\lim_{t_\mathrm{w}\to
    \infty} C(s t_{\mathrm w}^\mu,t_\mathrm{w})\,.$ The
translationally invariant sector is just $\mu=0$. In the range
$0<s<\infty$ the correlation function varies in
$q_\text{EA}<C_\infty^{(\mu=0)}(s)<1\,.$ Full Aging~\cite{RODRIGUEZ}
would imply that when $0<s<\infty$, $C_\infty^{(\mu=1)}(s)$ goes from
$q_\mathrm{EA}$ to 0. At the present moment, it is unclear whether
the full range of variation of the correlation function, $0<C<1$, may
be covered with just two time sectors.}
\begin{equation}\label{Cinf}
C_\infty(t)=\lim_{t_\mathrm{w}\to\infty} C(t,t_\mathrm{w})\,,
\end{equation}
as
\begin{equation}\label{EQ-QEA}
q_\mathrm{EA}=\lim_{t\to\infty} C_\mathrm{\infty}(t)\,.
\end{equation}
\end{enumerate}
The computation of $q_\mathrm{EA}$ is notoriously difficult~\cite{SUE2}.
Note that other authors~\cite{LET-NUM1} subtract $q_\text{EA}$
from $C_\infty(t)$ in such a way that it tends to zero
for large $t$.

\item The link  correlation function
\begin{equation}
C_\mathrm{link}(t,\tw)=\overline{\frac{1}{3N}\sum_{\langle \vn{x},\vn{y}\rangle}  c_{\vn x}(t,t_\mathrm{w}) c_{\vn y}(t,t_\mathrm{w})}\,,
\end{equation}
carries information on interfaces. Indeed, consider a coherent
spin-flip in a domain half of the system size. This will induce a
dramatic change in $C(t,t_\mathrm{w})$. On the other hand, the change in
$C_\mathrm{link}$ will be concentrated at the lattice links that are
cut by the surface of the flipped domain. If the geometry of this
flipped region is that of a compact object with a vanishing surface
to volume ratio, $C_\mathrm{link}$ will remain basically unchanged.\\

\end{itemize}

\item {\em Space} dependent correlation functions:\\
\begin{itemize}
\item Single time correlation function:
\begin{equation}
C_4(\vn r,\tw)=\overline{\frac{1}{N}\sum_{\vn x} q_{\vn x}(\tw) q_{\vn x +\vn r}(\tw)}\,.\label{C4-DEF}
\end{equation}
The long distance decay of $C_4(\vn r,\tw)$ defines the
coherence length:
\begin{equation}
 C_4(\vn r,\tw)\sim \frac{1}{r^a} f\left(r/\xi(t_\mathrm{w})\right)\,.\label{C4-DECAY}
\end{equation}
The exponent $a$ is relevant, because $C_4$ at distances
$\xi(t_\mathrm{w})$ tends to zero as $\xi(\tw)^{-a}$.  For
coarsening systems, because $a=0$, the order parameter does not vanish
inside a domain.  For the Ising SG in three dimensions the exponent
was found to be $a\approx 0.4$~\cite{DYNJANUS,GROUND} (see Sect.~\ref{SECT-a} for details).
The long distance damping
function $f$ seems to decay faster than exponentially,
$f(x)=\mathrm{exp} [-x^\beta]$ with $ \beta\sim 1.5\,$
\cite{SUE3,MPRTRL_1999}. Note as well that, at the critical point, $a$
is related to the anomalous dimension, the latest estimate being
$a(T_\text{c})=1+\eta=0.625(10)$~\cite{PELISSETTO}. The physical origin
for a nonzero $a$ below $T_\text{c}$ is in the replicon mode. In fact,
it was conjectured that for all $T<T_\text{c}$, $a(T) = a(T_\text{c})/2$~\cite{DeDominicis}.
In~\cite{DYNJANUS} we found values not far from this prediction. Note
that the exponent $a$ is discontinuous at $T_\text{c}$~\cite{PRRR}.

\item The two-time spatial correlation function (see~\cite{LET-NUM1,LET-NUM2})
\begin{equation}
C_{2+2}(\vn r,t,\tw)=\overline{\frac{1}{N}\sum_{\vn x} \left[c_{\vn x}(t,\tw) c_{\vn x +\vn r}(t,\tw)\ -\ C^2(t,\tw)\right]}\,,
\label{C22-DEF}
\end{equation}
(one could also subtract $\overline{C(t,\tw)}^2$, but due to the self-averaging
character of $C$ this leads to  the same thermodynamic limit).
This correlation function is rather natural for the structural glasses
problem, see for instance~\cite{BIROLI-BOUCHAUD}, where an adequate
order parameter is unknown.

There is a simple probabilistic interpretation of $C_{2+2}$. Let us
call a {\em defect} a site where $c_{\vn x}(t,\tw)=-1$ and
let $n(t,t_\mathrm{w})$ be the density of these defects. We trivially have
$C(t,t_\mathrm{w})= 1-2 n(t,t_\mathrm{w})$. The conditional
probability of having a defect at site $\vn x+\vn r$ knowing that
there already is a defect at site $\vn x$ is $n(t,t_\mathrm{w})g(\vn
r)$, where the defects' pair-correlation function is $g(\vn
r)$. Hence, $C_{2+2}(\vn r,t,\tw)$ is just
$4n^2(t,t_\mathrm{w})\bigl(g(\vn r) -1\bigr)$.

The long distance decay of $C_{2+2}(\vn r,t,\tw)$ defines the
correlation length $\zeta(t,t_\mathrm{w})$\footnote{%
The difference between coherence and correlation length is a subtle one.
In this work, we shall reserve the name `coherence length', which is computed
from a non-connected correlation function, Eq.~\eqref{C4-DEF}, for
the typical size of the coherent domains. The correlation length, which
is computed from a connected correlation function, Eq.~\eqref{C22-DEF},
refers to the characteristic length for defect correlation. In particular,
the coherence length diverges when $\tw\to\infty$, while the correlation
length may or may not diverge in that limit.}

\begin{equation}
C_{2+2}(\vn r,t,\tw)\sim \frac{1}{r^b} g\left(r/\zeta(t,t_\mathrm{w})\right)\,.\label{C22-DECAY}
\end{equation}
Basically nothing is known on exponent $b$ nor on the long distance
damping function~$g$. In~\cite{LET-NUM1,LET-NUM2}, this decay was fitted with $b=0$ and
$g(x)=\mathrm{e}^{-x}$, but the smallness of the found correlation lengths
$\zeta(t,t_\mathrm{w})<2$ for $t+t_\mathrm{w}\leq 1.3\times
10^8\,$~\cite{LET-NUM1,LET-NUM2}, does not permit strong claims. In the structural
glasses context~\cite{BIROLI-BOUCHAUD}, one tries to interpret
$\zeta(t,t_\mathrm{w})$ as a coherence length such as $\xi(t_\mathrm{w})$,
rather than as correlation length. As we shall empirically show, this might be
very reasonable in the limit $t\gg t_\mathrm{w}$.

In an RSB framework, the relaxation within a single state corresponds
to the range $q_\text{EA}<C(t,t_\mathrm{w})<1$ (the further decay of
$C(t,t_\mathrm{w})$ corresponds to the exploration of new
states). This regime is quite naturally identified with the
condition that  $\zeta(t,t_\mathrm{w})~\ll~\xi(t_\mathrm{w})$. In fact, $q_\mathrm{EA}$ yields the (correlated) percolation threshold for defects.

Sometimes we will find it useful to change variables from $t$ to
$C$. This is always feasible, because
$C(t,t_\mathrm{w})$ is a monotonically decreasing function of $t$ for fixed $t_\mathrm{w}$. The
accuracy of our numerical data allows this change of variable
without difficulty
(we have used a cubic spline, since the function $C(t,t_\mathrm{w})$ was sampled at a selected set of times).

Finally, note that $C_{2+2}$ is the difference of two statistically
correlated quantities (hence, the statistical error in the difference
may be expected to be smaller than that for each of the two
terms). This can be adequately taken into account by means of a jackknife
procedure (see, e.g.,~\cite{VICTORAMIT}).
\end{itemize}

\end{enumerate}

All the quantities defined so far are self-averaging (i.e., their
relative errors for a fixed number of samples
decrease as $N^{-1/2}$), with the notorious exception of
$\chi_\mathrm{SG}(t_\mathrm{w})$. This fact provides justification for
the standard strategy in nonequilibrium studies (both numerical and
experimental!)  of averaging results over very few samples.

Self-averageness stems from the fact that the computed/measured
quantities are averages of local observables taken over the full
system (which provides a number of statistically independent summands
of the order of $L^D/\xi^D(t_\mathrm{w})$. The exception,
$\chi_\mathrm{SG}(t_\mathrm{w})$ (\ref{CHISG-DEF}), is actually non local as it is the
integral over the whole system of $C_4(\vn r,t_\mathrm{w})$, Indeed,
the central limit theorem suggests that the probability distribution
function of $q(t_\mathrm{w})$ should tend to a Gaussian when
$L\to\infty$.  Hence, the variance for
$\chi_\mathrm{SG}(t_\mathrm{w})$ is $\sim 2
\chi^2_\mathrm{SG}(t_\mathrm{w})$ in the limit of a large system.

\subsection{Simulation details}\label{DESCRI-SIMU}

The \emph{Janus} computer~\cite{JANUS} can be programmed for the simulation
of the single spin flip Heat Bath dynamics up to a very large number of
lattice sweeps (units of Monte Carlo time) for systems of linear sizes up to
$L\sim100$.\footnote{The overall parallel update rate for $L=80$ systems
of the whole Janus ---$256$ nodes--- is 78 femtoseconds per spin.}

We have spent the most effort in simulating the dynamics of the
model described by Eq.~(\ref{EA-H}) in the direct quench protocol
described in the Introduction, for several runs of about
a hundred samples of linear size $L=80$ and up to $10^{11}$ Monte Carlo
steps (we recall that a single step corresponds to
roughly 1 picosecond of time in the real world). Details of
our simulations are given in Table~\ref{TAB-PARAMETERS}.
We extend here the analysis of the simulations reported on~\cite{DYNJANUS},
but additional simulations have also been carried out.
Most notably, we simulated $768$ new samples of size $L=80$ at $T=0.7$ up to $10^{10}$ in
Monte Carlo time, which have been useful to improve and test the statistical
accuracy in some aspect of our analysis.

We wrote to disk the spin configurations at
all times of the form $[2^{i/4}]+[2^{j/4}]$,
with integer $i$ and $j$ (the square brackets
stand for the integer part). Hence, our $t$ and
$\tw$ are of the form $[2^{i/4}]$. Nevertheless,
we computed $C_{2+2}$ only for powers of two,
due to the increased computational effort.

A final note on the time span of our runs. Much to our surprise,
we found in~\cite{DYNJANUS} that, even for our very large systems,
finite size effects in the coherence length
can be resolved with our statistical accuracy.
In this work, we have restricted ourselves to the time window
that is not affected by them. The single exception will
be in the analysis of energy relaxation, Sect.~\ref{SECT-ENERGY},
where this range is too short. Nevertheless, we have explicitly
checked that the energy suffers from smaller finite size
effects than the coherence length.
\begin{table}[hbtp]
\centering
\label{tab:simpar}
\begin{tabular*}{\columnwidth}{@{\extracolsep{\fill}}clcc}
\hline
$L$ &\multicolumn{1}{c}{$T$} & MC steps & $N_s$\\
\hline
$80$ & $0.6$   & $10^{11}$          & $96$ \\
$80$ & $0.7$   & $10^{11}$          & $63$ \\
80 &  0.8   & ${10}^{11}$          & 96 \\
\bfseries 80 & \bfseries 0.9   & $\textbf{2.8}\boldsymbol\times\textbf{10}^\textbf{10}$ & \bfseries 32 \\
$80$ & $1.1$   & $4.2\times 10^{9}$  & $32$ \\
\bfseries 80 & \bfseries 1.15  & $\textbf{2.8}\boldsymbol\times \textbf{10}^\textbf{10}$ & \bfseries 32 \\
\bfseries 80 & \bfseries 0.7   & $\textbf{10}^\textbf{10}$          & \bfseries 768 \\
$40$ & $0.8$   & $2.2\times 10^8$    & $2218$ \\
\hline
\end{tabular*}
\caption{Parameters of our simulations.
The overall \emph{wall-clock} time needed was less than six weeks.
We highlight with boldface the simulations performed after completion of~\cite{DYNJANUS}.
Recall that we take the critical temperature from~\cite{PELISSETTO}, $T_\text{c}=1.109(10)$.
The full analysis of spin configurations was performed offline.}\label{TAB-PARAMETERS}
\end{table}

\subsection{Fits for extremely correlated data}\label{DIAGONAL-CHI-TWO-SECT}
Computing the best fit parameters and estimating errors from
extremely correlated data sets presents a
common, and still not satisfactorily solved, difficulty
in many numerical studies. For instance, our study
of $C(t,\tw)$ requires considering approximately $10^4$
random variables extracted from a set of only $63$--$768$
samples. The standard approach, computing the covariance
matrix and inverting it, fails because this matrix is necessarily
singular.\footnote{Consider computing the covariance
matrix for a set of $N_O$ random variables from
$N_\text{s}< N_O$ samples. If the $N_\text{s}\times N_O$
numbers are disposed on a rectangular matrix,
it is clear that the last $N_O-N_\text{s}$ rows (reordering the rows if necessary)
are a linear combination of the others. In other words,
the results are indistinguishable from the ill-conditioned
situation where these last $N_O-N_\text{s}$ random variables
are the very same linear combination of the first ones. Once
this is realized it is trivial to show that the range of the
size $N_O\times N_O$ covariance matrix is at most $N_\text{s}$.}
In this paper we shall follow an empirical procedure.
We shall consider only the diagonal part of the covariance
matrix in order to minimize $\chi^2$ when performing fits.
Unless otherwise indicated, we shall always use this diagonal $\chi^2$ in the rest
of the paper.
Yet, in order to take correlations into account we shall
perform this procedure for each jackknife block and
later on compute error estimates from their fluctuations.
As we have run simulations with both 63~(from \cite{DYNJANUS}) and 768 samples
for $T=0.7$, we are in a position to test this method
by comparing the results obtained and the ones to be expected for 63 samples
(see Sect.~\ref{SECT-a}).

The main drawback of this approach is that
the standard $\chi^2$ test of fit likelihood cannot be applied blindly.
Of course, were the exact fitting function known, the average value
of diagonal $\chi^2$ should be $1$ per degree of freedom. Yet, since
the obtained fitting function may coherently fluctuate with the numerical data,
we shall encounter anomalously low values of  diagonal $\chi^2$.  We examine
this problem in Sect.~\ref{SECT-a}, empirically finding that $\chi^2$ behaves
as if there were many fewer degrees of freedom than what one would expect.

\section{Integral estimators of characteristic length scales}\label{XI-SECT}
The need to estimate characteristic length scales, such as
$\xi(t_\mathrm{w})$ or $\zeta(t,\tw)$
is a recurrent theme in lattice
gauge theory and statistical mechanics. The more straightforward
method is to consider a particular functional form for the long
distance damping function in Eqs.~(\ref{C4-DECAY}) or (\ref{C22-DECAY}).
One of the problems with this approach, already identified in the
study of equilibrium critical phenomena, is that it is extremely
difficult to extract from numerical data {\em simultaneously} the
length scale and the exponent for the algebraic decay. Note that quite
often computing the exponent is as important as extracting the length
scale to draw physical conclusions. The situation worsens if the
functional form is only an educated guess, which
is precisely our case. Furthermore, numerical data for the correlation
function at different lattice sites  suffer from  dramatic
statistical correlations, which complicates fitting
procedures.

A different approach, the use of integral estimators,
has been known since the 1980s~\cite{COOPER}; but only
in the mid 1990s (see e.g.~\cite{FSS-PISA,QUOTIENTS}) it was realized
that it provided an enormous simplification. The use of integral estimators for
the length scale enables determinations of exponents such as $a$
in Eq.~(\ref{C4-DECAY}), which are completely independent from the
functional form of the long distance damping. The only place left for
systematic errors is in finite size effects or in scaling corrections
(when the considered range for the variation of length scales such as
$\xi(\tw)$ is too small). As for the determination of the length scale
itself, integral estimators are guaranteed to produce numbers that
scale as the inaccessible {\em true} $\xi(\tw)$, provided that it is
large enough.

The fact that the correlation functions that will be considered here,
$C_4$ and $C_{2+2}$, are self-averaging in a nonequilibrium context
provides an impressive error reduction, which is not accessible for
equilibrium studies.

Our chosen example to explain the method will be that of $C_4$ and
the determination of the coherence length and of the exponent $a$.

\subsection{The coherence length}
Cooper {\em et al.\/}~\cite{COOPER} suggested the second moment
determination of the characteristic length,
\begin{equation}\label{SECOND-MOMENT}
\xi^{(2)}(\tw) \equiv \frac{1}{\sqrt2 \sin \pi/L}\left[ \frac{\hat C_4(0,\tw)}{\hat C_4(\vn k_\text{min},\tw)} -1 \right]^{1/2},
\end{equation}
where $\hat C_4(\vn k,\tw)$ is the Fourier transform of $C_4(\vn r,\tw)$, and
$\vn k_{\text{min}}$ is the minimal non-vanishing wave vector allowed by
boundary conditions ($\vn k_{\text{min}}=(2\pi/L,0,0)$ or permutations).
Notice that $\chi_\text{SG}(\tw)=\hat C_4(0,\tw)$.
As can be readily seen, in the thermodynamic limit this is equivalent to
\begin{equation}
\xi^{(2)}_{L=\infty}(\tw) = \sqrt{\frac{\int\mathrm{d}^D\vn r\ \vn r ^2 C_4(\vn r,\tw)}{\int\mathrm{d}^D \vn r \ C_4(\vn r,\tw)}}.
\end{equation}
The denominator in this equation is just the SG susceptibility~\eqref{CHISG-DEF} which,
as we said in  Sect.~\ref{SECT-OBSERVABLES}, does not self-average (and neither does
the numerator). Because of this, if one were to follow this method, a very large
number of samples would be needed.

We would like to have a better statistically behaved definition
of $\xi(\tw)$. In order to get it, we start by considering the integrals\footnote{In what follows
$C_4(r,\tw)$ stands for $C_4(\vn r,\tw)$, with $\vn r  = (r,0,0)$
and permutations. As we shall see in Sect.~\ref{SECT-ISOTROPY},
using an average over spherical shells does not achieve a significant reduction
of statistical errors in our chosen estimator for the coherence length.}
\begin{equation}\label{Ik}
I_k(\tw)\equiv\int_0^{L/2} \mathrm{d} r\ r^k C_4(r,\tw).
\end{equation}
As we are going to work in the thermodynamical limit, we are interested
in the regime $L\gg \xi(\tw)$, so we can safely reduce the upper integration
limit from $\infty$ to $L/2$.

\begin{figure}[t]
\centering
\includegraphics[height=\linewidth,angle=270]{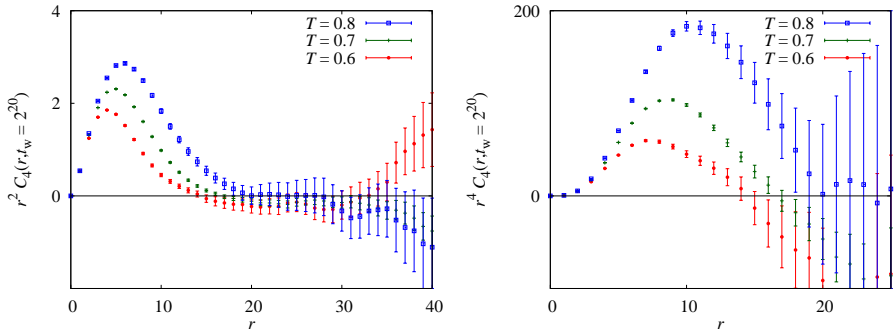}
\caption{The spatial autocorrelation of the overlap field
for $\tw=2^{20}$ and three subcritical temperatures, as
computed in our $L=80$ lattice.
On the left panel we show $r^2 C_4(r,\tw)$, recall~\eqref{Ik},
while on the right one we show $r^4 C_4(r,\tw)$ (mind
the different scales). While the signal to noise ratio
of both quantities falls equally rapidly, the problem
is less severe for the computation of $I_2(\tw)$ since
the maximum there is not in the noise dominated region.
The curve for $T=0.7$ is the average of 768 samples,
while those of $T=0.6$ and $T=0.8$  are computed from
96 samples.}\label{CUTOFF}
\end{figure}

With this notation and assuming rotational invariance,
the second moment coherence length is just
\begin{equation}\label{SECOND-MOMENT-Ik}
\xi^{(2)}_{L=\infty}(\tw) \simeq \sqrt{\frac{I_{D+1}(\tw)}{I_{D-1}(\tw)}}\ .
\end{equation}
We also recall that in~\cite{KISKER} it was proposed to identify
$\xi(\tw)$ with $I_0(\tw)$,\footnote{The authors of~\cite{KISKER}
carefully discussed the interplay of the integration limits in~\eqref{Ik}
and the boundary conditions. Since we restrict ourselves to $L\gg \xi(\tw)$,
this is immaterial to us.} but this would only be
appropriate for $a=0$. For a correlation function
following the scaling law~\eqref{C4-DECAY}, one can use
a more general definition, because $I_k(\tw)\propto \left[\xi(\tw)\right]^{k+1-a}$:
\begin{equation}\label{XI-INTEGRALES}
\xi_{k,k+1}(\tw) \equiv \frac{I_{k+1}(\tw)}{I_k(\tw)} \propto \xi(\tw).
\end{equation}
Definitions such as \eqref{SECOND-MOMENT} and \eqref{XI-INTEGRALES}
suffer from systematic errors because equation~\eqref{C4-DECAY} is only
an asymptotic formula for large values of $r$. Therefore,
the systematic errors
in these definitions can be reduced by considering a large value
of $k$ (since the $r^k$ factor would suppress the deviations at short
distances). However, there is also the issue of statistical errors
to consider. As we can see in Fig.~\ref{CUTOFF}, a large
power of $r$ pushes the maximum of $r^k C_4(r,\tw)$ into
the region where $r\gg \xi(\tw)$ and the signal to noise
ratio of the correlation function is extremely low.
Because of this, a compromise in the choice of $k$ is needed.
Our preferred option is $\xi_{1,2}(\tw)$.

\begin{figure}
\centering
\includegraphics[height=\linewidth,angle=270]{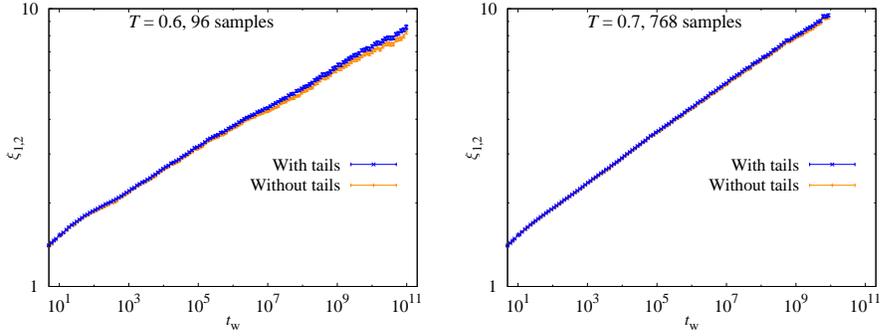}
\caption{\textbf{Left:} Result of computing
$\xi_{1,2}$ in two different ways for our 96 samples at $T=0.6$.
 In the orange curve (color online) we stop the integration at
the cutoff point where relative error of $C_4$ is greater than one third. In the blue
curve (color online) we estimate the contribution of the tail from that point on extrapolating with
a fit to~\eqref{XI-ALPHA-BETA}. The difference is small, but with the second method
the power law behavior of $\xi_{1,2}(\tw)$ lasts longer.
\textbf{Right:} Same plot for our 768 samples at $T=0.7$. With the increased
statistics this extrapolation is not as important and both
curves are compatible for the whole simulation. With
the 63 samples of~\cite{DYNJANUS}, the tail contribution
is as significant as in the left panel.}
\label{COLAS}
\end{figure}

Even though our use of $\xi_{1,2}(\tw)$ instead of $\xi^{(2)}(\tw)$ already
mitigates the statistical problems in the integration
of $C_4(r,\tw)$, we can still improve the computation.
As can be plainly seen in Fig.~\ref{CUTOFF}, $r^2 C_4(r,\tw)$
starts having very large fluctuations only at its tails,
where the contribution to the integral is minimal.
To take advantage of this fact, we are going
to use a self-consistent integration cutoff
(a method applied before in the study of correlated time series~\cite{SOKAL}).
We only integrate our data\footnote{%
As our (somewhat arbitrary, yet irrelevant) choice of quadrature method we have
chosen to interpolate the data with a cubic spline, whose integral
can be exactly computed.}
for $C_4(r,\tw)$ up to the point where this function first becomes
less than three times its own statistical error.
Of course, while this method provides a great reduction in statistical
errors, it does introduce a systematic one. To avoid it, we estimate
the small contribution of the tail with a fit to
\begin{equation}\label{XI-ALPHA-BETA}
C_4(r,\tw) = \frac{A}{r^{0.4}} \exp\left[ -\bigl(r/\xi^\text{fit}(\tw)\bigr)^{1.5}\right]\, .
\end{equation}
Notice that this is just the scaling function~\eqref{C4-DECAY},
using $f(x)=\exp[-x^{1.5}]$ as our damping function and $a=0.4$. Of course,
while this fit is used to estimate the contribution of the interval
$[r_\text{cutoff}, L/2]$, we actually perform the fitting for
$3\leq r\leq \min \{15,r_\text{cutoff}\}$, where
the signal is still good.  This
last step is important for large $\tw$ (see Fig.~\ref{COLAS}).
\begin{figure}[h]
\centering
\includegraphics[height=.7\linewidth,angle=270]{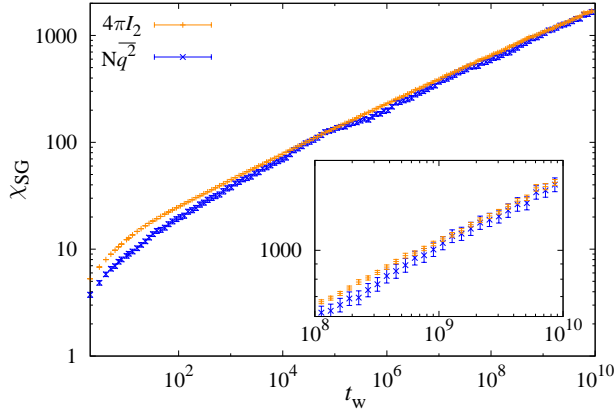}
\caption{The SG susceptibility $\chi_{\mathrm{SG}}(\tw)$ for our 768 samples at
$T=0.7$ computed
from $q^2$ and from the integral $I_2$: $\chi_{\mathrm{SG}}(\tw)=
N\overline{q^2(\tw)} = 4\pi I_2(\tw)$. The main difference between the two
determinations is that the second one has been computed
with a self-consistent cutoff. As we can see, even though
both curves are compatible, the integral one is much more precise.
The inset details the upper right corner.}
\label{SUSCEPTIBILITY}
\end{figure}

As a consistency check of this method and as a  demonstration
of its enhanced precision we can consider the SG susceptibility~\eqref{CHISG-DEF}.
This observable, $\chi_{\mathrm{SG}}(\tw)=N \overline{q^2(\tw)}$,
coincides with $4\pi I_2(\tw)$ in the presence of rotational invariance.
We have plotted both expressions as a function of time in Fig.~\ref{SUSCEPTIBILITY}.
The only systematic discrepancy between the two is at short
times, when the system cannot be considered rotationally invariant
(see Sect.~\ref{SECT-ISOTROPY}). However,
the integral determination $4\pi I_2(\tw)$ is much more precise 
for the whole span of our simulation.

\begin{figure}
\centering
\includegraphics[height=0.7\linewidth,angle=270]{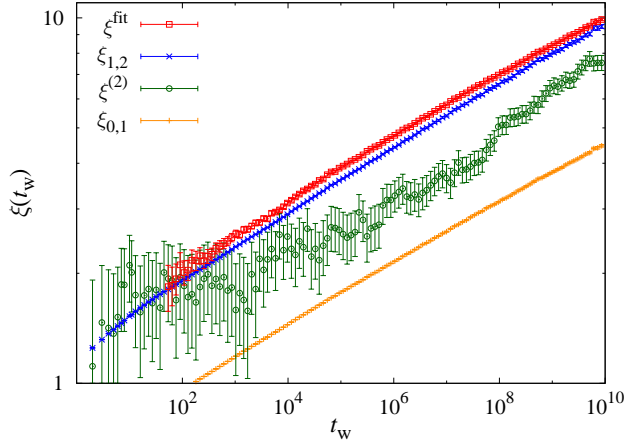}
\caption{Comparison of our integral estimators
$\xi_{0,1}$ and $\xi_{1,2}$, Eq.~\eqref{XI-INTEGRALES}, with the
second moment estimate $\xi^{(2)}$ and the result $\xi^\text{fit}$ of a fit to~\eqref{XI-ALPHA-BETA},
with $a=0.4$. All the curves become parallel at large $\tw$, but the integral
estimators have much smaller errors. All curves are for our 768 samples
at $T=0.7$.}
\label{METODOS-XI}
\end{figure}

As a second check, we have plotted in Fig.~\ref{METODOS-XI} the integral estimators $\xi_{0,1}(\tw)$ and $\xi_{1,2}(\tw)$,
together with the traditional second moment estimate $\xi^{(2)}$
and the result of a fit to \eqref{XI-ALPHA-BETA}. As we can see,
all determinations are indeed proportional, but the integral estimators
are much more precise.

\begin{figure}
\centering
\includegraphics[height=\linewidth,angle=270]{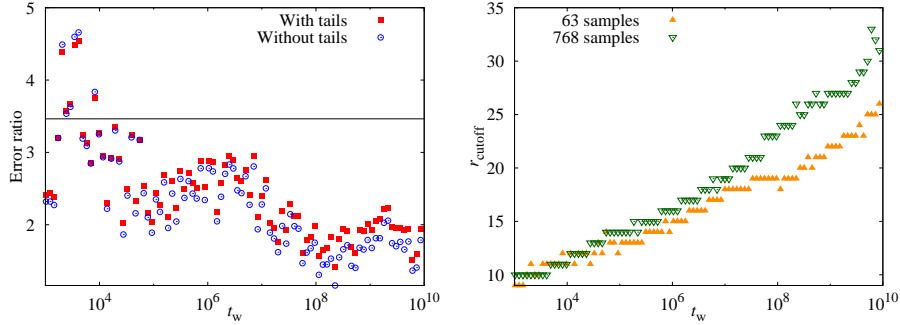}
\caption{
\textbf{Left:} Ratio of the errors in $\xi_{1,2}(\tw)$ for the simulations
at $T=0.7$ with the 63 samples of~\cite{DYNJANUS} and those of
our simulations with 768 new samples (see text for discussion).
The extrapolation to include the tails in the integrals
is immaterial for this ratio. The horizontal
line is $\sqrt{768/63}\approx3.46$. \textbf{Right:}
Cutoff of the $I_k$ integrals as a function of time
for both simulations.}\label{ERRORES-63-768}
\end{figure}
\begin{figure}
\centering
\includegraphics[height=.6\linewidth,angle=270]{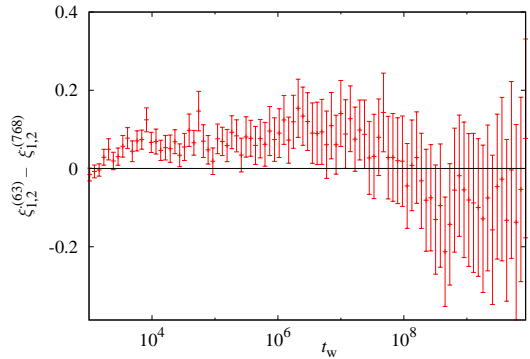}
\caption{
Difference between the coherence length $\xi_{1,2}(\tw)$
computed with the 63 samples of~\cite{DYNJANUS}
and with the 768 new samples (the errors
are the quadratic sum of those for each simulation).
Both curves are
compatible in the whole time range. Mind
the dramatic statistical correlation in the sign
of this difference.}\label{XI-63-768}
\end{figure}

Finally, we address the issue of our error estimates
by comparing our data at $T=0.7$ for the 63 samples of~\cite{DYNJANUS}
with our new simulations with 768 samples.
We have explicitly checked that the errors
in $C_4(r,\tw)$, computed with the jackknife method,
scale as the inverse of the square root of the number of samples,
within errors (for Gaussian distributed data, the relative statistical
error in the error estimate is $\sim 1/\sqrt{2 N_\text{samples}}$).
The fact that $C_4$ verifies this basic expectation is
a demonstration that large deviations, which would be missed
for a small number of samples, do not appear, even for
$\tw$ as large as $10^{10}$. On the other hand,
the behavior of statistical errors in
integrals with a dynamically fixed cutoff
is not that simple~\cite{SOKAL}.
In our case, the ratio of the errors
in $\xi_{1,2}(\tw)$ (Fig.~\ref{ERRORES-63-768}, left)
is around $30\%$ below the Gaussian
expectation. We have checked that a similar
effect arises in the computation of $I_2(\tw)$
and that the effect of the tails is immaterial.
In fact, this deviation is entirely due to the difference
in the dynamical cutoffs of both simulations, Fig.~\ref{ERRORES-63-768},
right. Whenever the cutoff coincides, the error ratio is
in the expected region around $\sqrt{768/63}\approx3.46$.
The overall consistency of our error determinations
is demonstrated by Fig.~\ref{XI-63-768}.

\subsection{The algebraic prefactor}\label{SECT-a}
One of our main goals
is to provide a precise estimate of the exponent $a$ for
the algebraic part of $C_4$, see equation~\eqref{C4-DECAY}.
Equilibrium methods~\cite{QUOTIENTS} are not well suited
to a nonequilibrium study in the thermodynamic limit.
Instead, we introduce here the method used, but not
explained, in~\cite{DYNJANUS}.

The starting point is the realization that $I_1\propto \xi_{1,2}^{2-a}$,
which would indicate that $a$ can be obtained from a power law
fit of $I_1$ as a function of the coherence length. Furthermore,
the large statistical correlation between $I_1$ and $\xi_{1,2}$
can be used to reduce the statistical errors in $a$. However, such a fit
would be quite problematic, as we would have errors on both coordinates
(the correlation of the data already poses a nontrivial problem
with errors in just one coordinate, see Sect.~\ref{DIAGONAL-CHI-TWO-SECT}).
Instead, we fit separately
$I_1(\tw)$ and $\xi_{1,2}(\tw)$ to power laws in the waiting time.
This way, if
\begin{align}
I_1(\tw) &= A \tw^{c},&
\xi(\tw)& =B \tw^{1/z},\label{XI-GROWTH}
\end{align}
we have  $a=2-cz$. This relation holds for each jackknife block,
which lets us take full advantage of the correlations.

Following this method, we obtain the results in Table~\ref{TAB-a}.
In~\cite{DYNJANUS} (first four rows of Table~\ref{TAB-a})
we quoted the results for a fitting
range of $\xi_{1,2}\in[3,10]$, which is perfectly adequate
for $T=0.6,0.8$ and $1.1$. As we can see, $\chi^2$ appears to be a bit too large
for $T=0.7$, but if we narrow the fitting range it becomes
reasonable and $a$ does not change.

As this is a very important magnitude, in this work
we have increased the number of samples
at $T=0.7$ by a factor of $12$ with respect to~\cite{DYNJANUS},
(from $63$ to $768$ samples, with extra simulations that stop at $\tw=10^{10}$,
rather than $10^{11}$). This not only allows us to provide a better
estimate of $a$ but we are now also able to check the soundness
of the statistical procedure.

The first difficulty is that,
with the corresponding reduction in statistical errors, the
original fitting window no longer provides reasonable values
of our diagonal $\chi^2$ estimator.
Instead, we have pushed the lower limit to $\xi\geq 4$ (see Table~\ref{TAB-a} for
details). The new value
of $a(T=0.7)$ is
\begin{equation}
a(T=0.7) = 0.397(12).
\end{equation}
\begin{table}
\centering
\begin{tabular*}{\columnwidth}{@{\extracolsep{\fill}}clcrlcc}
\hline
$T$ &
$N_\text{samples}$&
$[\xi_\text{min},\xi_\text{max}]$&
\multicolumn{1}{c}{ $z$} &
\multicolumn{1}{c}{ $a$} & $\chi^2_{\xi}$/d.o.f. & $\chi^2_{I_1}$/d.o.f.\\
\hline
\multirow{1}{0.5cm}{$0.6$}
&\multirow{1}{0.5cm}{96}
 & $[3,10]$ & 14.06(25) & 0.359(13) & 41.7/82 & 49.0/82\\
\hline
\multirow{2}{0.5cm}{$0.7$}
&\multirow{2}{0.5cm}{63}
  &  $[3,10]$  & 11.84(22) & 0.355(15) & 82.7/81 & 131/81  \\
& &  $[3.5,10]$& 12.03(27)   & 0.355(17) & 52.7/71 & 75.5/81  \\
\hline
\multirow{1}{0.5cm}{$0.8$}
&\multirow{1}{0.5cm}{96}
 & $[3,10]$   & 9.42(15)  & 0.442(11) & 17.1/63 & 12.2/63\\
\hline
\multirow{1}{0.5cm}{$1.1$}
&\multirow{1}{0.5cm}{32}
 & $[3,10]$   & 6.86(16)  & 0.585(12) & 18.7/46 & 26.1/46\\
\hline
\multirow{4}{0.5cm}{$0.7$}
&\multirow{4}{0.5cm}{768}
  & $[3,10]$   & 11.45(10) & 0.395(8)  & 86.9/76 & 269/76\\
&  & $[3.5,10]$ & 11.56(13) & 0.397(10) & 46.6/66 & 101/66\\
&  & $[4,10]$   & 11.64(15) & 0.397(12) & 40.1/58 & 60.4/58\\
&  & $[4.5,10]$ & 11.75(20) & 0.394(14) & 29.6/50 & 35.8/50\\
\hline
\end{tabular*}
\caption{Value of the dynamic exponent $z$ and the algebraic prefactor
$a$ for several temperatures. The fitting range $\xi_{1,2}\in[3,10]$,
which worked for the smaller number of samples we had in~\cite{DYNJANUS},
does not give good fits for $I_1$ with our enlarged statistics at $T=0.7$
(the fits for the coherence length itself are still good). Nevertheless, if
we increase $\xi_\text{min}$ to get reasonable values of $\chi^2_{I_1}$/d.o.f.
we see that the estimate of $a$ does not change.}
\label{TAB-a}
\end{table}

\begin{figure}
\centering
\includegraphics[height=\linewidth,angle=270]{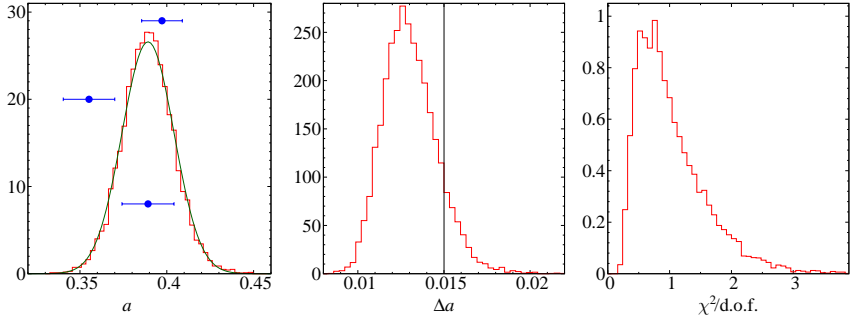}
\caption{\textbf{Left:} Probability density function $\rho(a)$
of the estimate of exponent $a$, Eq.~\eqref{C4-DECAY},
as obtained from a set of 63 samples. The dots with
\emph{horizontal} error bars are, from top to bottom:
our best estimate with 768 samples, the value
with the 63 samples of~\cite{DYNJANUS},
and the mean and standard deviation of $\rho(a)$.
The continuous line is a Gaussian distribution
with the same mean and variance as $\rho(a)$.
\textbf{Center:} As in left panel, for the jackknife
errors $\upDelta a$. The vertical line marks
the standard deviation of $\rho(a)$.
\textbf{Right:} Histogram of the $\chi^2$/d.o.f.
parameter for the $xi$ fit. In all three panels the fitting range to obtain $a$ was taken as
$\xi\in[3,10]$.} \label{HISTOGRAMA-a}
\end{figure}
In accordance to the analysis of Fig.~\ref{ERRORES-63-768},
the error has not decreased the factor $\sqrt{768/63}$
with the increase in statistics. One could say that the
dynamic cutoff procedure has traded statistical uncertainty
for a reduction in systematic errors.  Another contributing
factor to the large statistical error is the raising
of the minimal coherence length included in the fit.

To understand whether the discrepancy between the
new estimate of $a$ and the one in~\cite{DYNJANUS}
is due to a systematic effect or to a large fluctuation,
we can use a Monte Carlo method. The probability
distribution function of the estimates of $a$,
as computed with 63 samples, can be obtained easily
from our set of 768 samples. One randomly picks 
sets of 63 different samples and determines $a$ and
its jackknife error $\upDelta a$ for each of these sets
(mind that there are $\binom{768}{63}\approx2.2\times 10^{93}$
possible combinations). We have done this 10\,000 times
and computed normalized histograms of both quantities (Fig.~\ref{HISTOGRAMA-a}).
Clearly enough, the estimate in~\cite{DYNJANUS} was a fluctuation
of size $2.2$ standard deviations, large but not unbelievably so.
On the other hand we see
that the jackknife method tends to slightly underestimate $\upDelta a$
for 63 samples (Fig.~\ref{HISTOGRAMA-a}, center).
Note as well that there seems to be a small bias (smaller than the error)
on the estimate of $a$ with only 63 samples (Fig.~\ref{HISTOGRAMA-a}, left,
compare the histogram with the uppermost horizontal point). There
are two possible reasons for this. One is that $a$ is obtained from
raw data through a nonlinear operation. The other is that
the larger the number of samples, the smaller the cutoff
effects in the computation of $I_k(\tw)$.

It is amusing to compute as well the probability density of
$\chi^2$/d.o.f. for fits with 63 samples. We show in Fig.~\ref{HISTOGRAMA-a}, right,
that $\chi^2_\xi$ for the fit of the coherence length
can be much larger than what one would naively expect for
a fit with $\sim80$ degrees of freedom.

\subsection{Isotropy of $C_4(\vn r,\tw)$}\label{SECT-ISOTROPY}
\begin{figure}[b]
\centering
\includegraphics[height=0.45\linewidth]{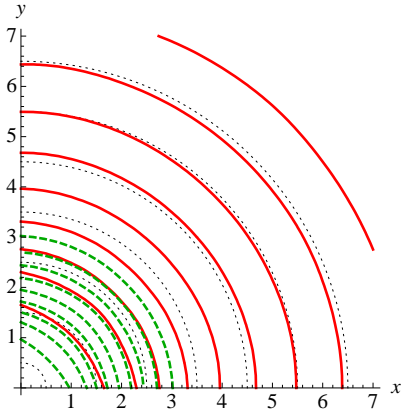}
\caption{Level curves $C_4(\vn r, \tw)=c$ for $c=0.3$ (dashed lines) and $c=0.1$ (solid lines)
at $T=0.6$ (dotted lines are circles, for visual reference). We have
restricted ourselves to the $z=0$ plane, for clarity. The innermost
curve corresponds in both cases to $\tw=4$ and the succeeding ones
correspond to geometrically growing times ($\tw=4\times 16^{i}$).
As we can see, the deviations from isotropy are mainly due to
lattice discretization (i.e., functions of $r$), even though there is also a small
dependence on time for curves of similar radii. The errors are smaller
than the thickness of the lines (the interpolation to draw these continuous
curves from our lattice data was performed with {\em Mathematica TM}).}
\label{ISOTROPY}
\end{figure}
In the previous section we neglected the issue of the isotropy of $C_4(\vn r,\tw)$.
At all times we worked with radial functions $C_4(r,\tw)$, obtained by averaging
the correlation at distance $r$ along the three axes. In doing this, we ignored
most of the $N$ points that $C_4(\vn r,\tw)$ has for a given $\tw$. The main
motivation for doing this is avoiding the computation of the whole correlation
function, a task which in a naive implementation is $\mathcal O(N^2)$. Of course,
due to the Wiener-Khinchin theorem, we can reduce this computation to the evaluation
of two Fourier transforms which, using an implementation of the FFT algorithm~\cite{FFTW},
is an $\mathcal O(N\log N)$ task.

We shall examine in this section  whether the complete correlation functions are
isotropic
and whether we can take advantage of them to reduce the errors in our determination
of the coherence length. The first question is answered by Fig.~\ref{ISOTROPY}, where
we compute the level curves $C_4(\vn r,\tw)= c$ for several values of $c$ and $\tw$.
As we can see, isotropy is recovered at quite small distances (remember we are only concerned
with $\xi\gtrsim 3$).

\begin{figure}
\centering
\includegraphics[height=\linewidth,angle=270]{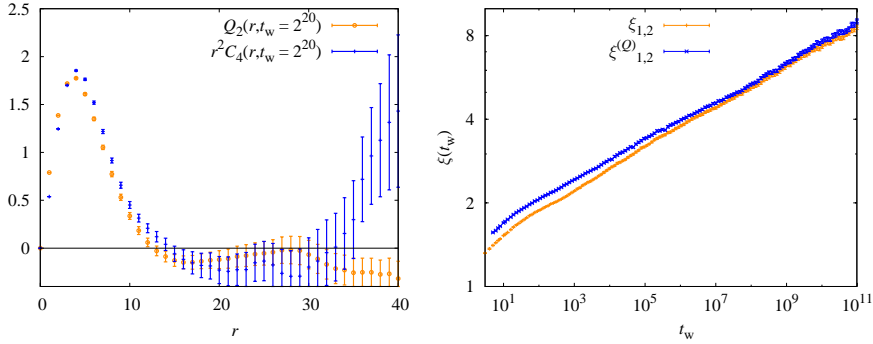}
\caption{\textbf{Left:} Comparison of $Q_2(r,\tw=2^{20})$
with $r^2 C_4(r,\tw=2^{20})$ for $T=0.6$. The first quantity is obtained
by averaging over spherical shells, while the second one considers
only correlations along the axes. The behavior of $Q_2$ is better
at the tails, but does not imply any gain in practice, as both
functions are equally well behaved up to the cutoff point. \textbf{Right:} The coherence length
computed with the whole correlation functions, $\xi^{(Q)}_{1,2}$, and with correlations
along the axes, $\xi_{1,2}$. Both estimates coincide for large
times.}
\label{XI-ISOTROPA}
\end{figure}

In order to use our integral method for a three-dimensional $C_4(\vn r,\tw)$ we must first
average it over spherical shells. We do this by defining the functions $Q_k(n,\tw)$,
\begin{equation}\label{Qk}
Q_k(n,\tw) \equiv \frac{\displaystyle \sum_{|\vn r| \in [n,n+1)} |\vn r|^k C_4(\vn r,\tw)}
{\displaystyle \sum_{|\vn r| \in [n,n+1)} 1}\ .
\end{equation}
Notice that $Q_0(0,\tw)=C_4(0,\tw)=1$ and that the division by the number of points
is needed to average over the spherical shell. Now we can use $Q_k(r,\tw)$ in the same way we used
$r^k C_4(r,\tw)$ in the previous section. The resulting
coherence length $\xi^{(Q)}_{1,2}(\tw)$ would be expected to coincide
with $\xi_{1,2}(\tw)$ in the large $\tw$ limit, but have much smaller errors
due to the large increase in statistics. As we can see from Fig.~\ref{XI-ISOTROPA},
however, the correlation among the points is so great that the gain in precision
is insignificant.

We can conclude from this section that the usual approximation of
considering $C_4(\vn r,\tw)$ isotropic is a well founded one and
that it is safe, and statistically almost costless,
to restrict ourselves to correlations along the axes, as
we did in~\cite{DYNJANUS} and in the previous section.
Nevertheless, the computation of the
whole $C_4(\vn r,\tw)$ could be rewarding if one were
to study $I_k$, with $k>2$.

\section{Characteristic length scales for dynamical heterogeneities}\label{XIC2TW-SECT}
\begin{figure}[t]
\centering
\includegraphics[height=\columnwidth,angle=270]{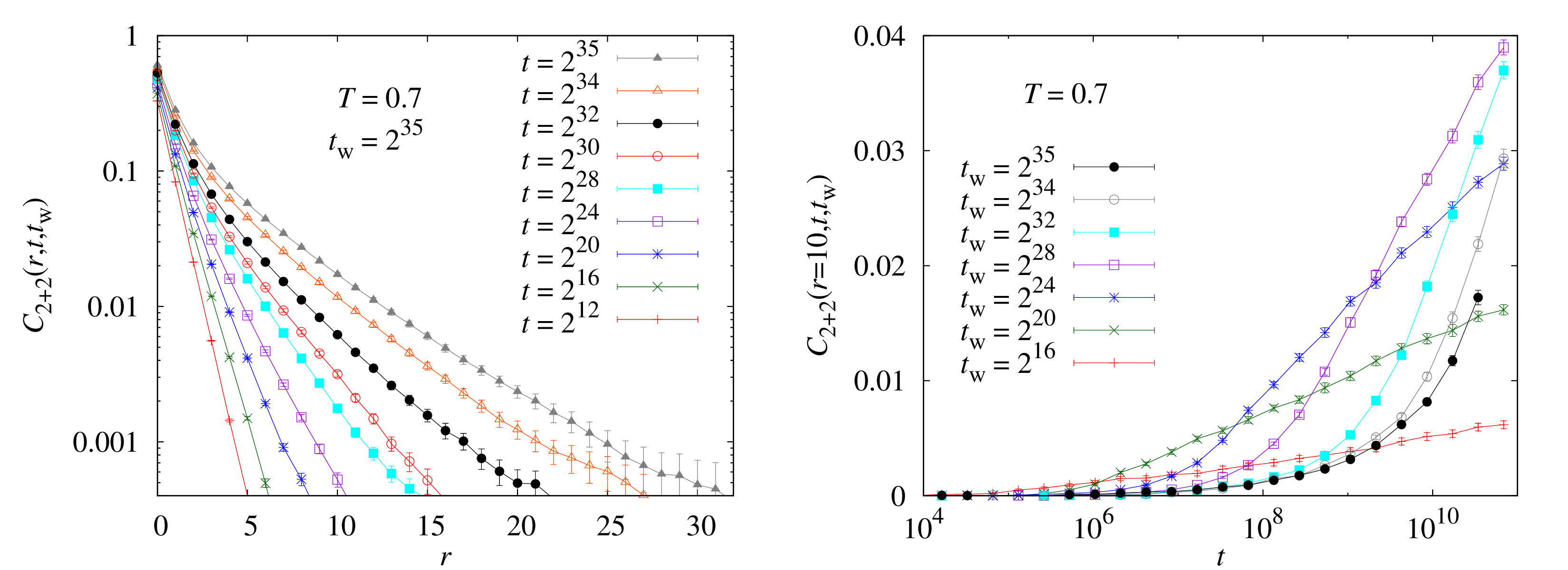}
\caption{{\bf Left}: The two-time spatial correlation function
  $C_{2+2}(r,t,\tw)$ (see Eq.~(\ref{C22-DEF})) as a function of $r$, for a high
  value of $\tw=2^{35}$, and several values of $t$. {\bf Right}:
  $C_{2+2}(r,t,\tw)$ as a function of $t$ and several values of $\tw$,
  fixing $r=10$. Both figures are for our 63 sample
  simulation at $T=0.7$.}
\label{fig:c22_r10_tw_t_0.7}
\end{figure}

\begin{figure}
\begin{center}
\includegraphics[height=\textwidth,angle=270]{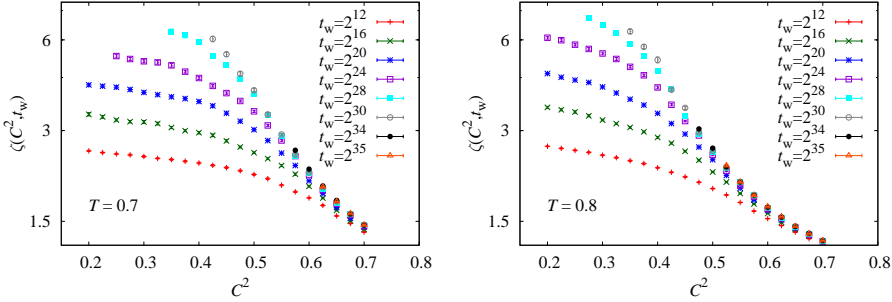}
\caption{{\bf Left}: Coherence length $\zeta(C^2,\tw)$ as a  function of $C^2$ for
  several values of $\tw$ at $T=0.7$ (63 samples). {\bf Right}: As figure
  left, for $T=0.8$ (96 samples).}
\label{fig:xiC2tw.vs.C2}
\end{center}
\end{figure}

The concept of heterogeneous dynamics has been recently borrowed from
structural glasses to describe non-local aging in Spin
Glasses~\cite{LET-NUM1,LET-NUM2}. A \emph{coarse-grained} spin correlation
function may be defined for which the microscopic average is not performed on
the overall volume but on cells of size $\ell^D$. Non-uniform aging of the
system results in spatial fluctuations of the coarse-grained correlation
functions, that can be used to define a two-time dependent correlation length.
In fact, at fixed system size and temperature one may study the
\emph{two-time} dependent distribution of the coarse-grained correlation
function values or, equivalently, the distribution as a function of $\tw$ and of
the global spin-spin correlation function $C(t,\tw)$,
Eq.~(\ref{SPIN-SPIN-CORRELATION}): if the coarse-graining size $\ell^D$ is larger than
the correlation length, then one should observe Gaussian statistics, while
strong deviations from a Gaussian distribution are present in case of small
$\ell^D$~\cite{LET-NUM2}. Such dependence of the statistics on the
coarse-graining size defines a crossover length $\zeta(C^2,\tw)$ interpreted
as an aging (\emph{two-time}) correlation length.

It has been observed~\cite{LET-NUM2} that such an aging correlation length may be
obtained from the spatial decay of the \emph{two-time} \emph{two-site}
correlation function $C_{2+2}$, Eq.~(\ref{C22-DEF}); still, the authors of
reference~\cite{LET-NUM2} could not measure $\zeta$ values greater than two
lattice spacings.  In this section we present data from our simulations on
Janus showing correlation lengths for dynamical heterogeneities up to order
ten lattice units.  We show $C_{2+2}(r,t,\tw)$ for $\tw=2^{35}$ and
some values of $t$ at temperature $T=0.7$ in the left picture of
Fig.~\ref{fig:c22_r10_tw_t_0.7} (as for $C_4$,
we denote by $C_{2+2}(r,t,\tw)$ the correlations along
the axes). As one can see in this figure, at large
times, correlations grow up to several lattice spacings.

In what follows it is convenient to eliminate the time $t$ dependence in favor
of the global spin-spin correlation function $C(t,\tw)$.  For given $\tw$ and
$C$ values, we obtain easily $t(\tw,C)$ because the monotonic time dependence
of the (discrete measures of the) correlation function can be smoothly
interpolated by means of cubic splines. Then, for each value of $r$, we
must interpolate $C_{2+2}(r,t(C,\tw),\tw)$. As one can see in the right
picture of Fig.~\ref{fig:c22_r10_tw_t_0.7} there is a sharp change in the
$t$-derivative of $C_{2+2}$, so we had to resort to a linear
interpolation in order to avoid the strong oscillations that a spline had
suffered from. Once $C_{2+2}(r,t(C,\tw),\tw)$ has been interpolated
at all $r$ values, the methods of Sect.~\ref{XI-SECT} allow us to estimate
the correlation length $\zeta(C^2,\tw)$.

\begin{figure}[t]
\begin{center}
\includegraphics[height=\textwidth,angle=270]{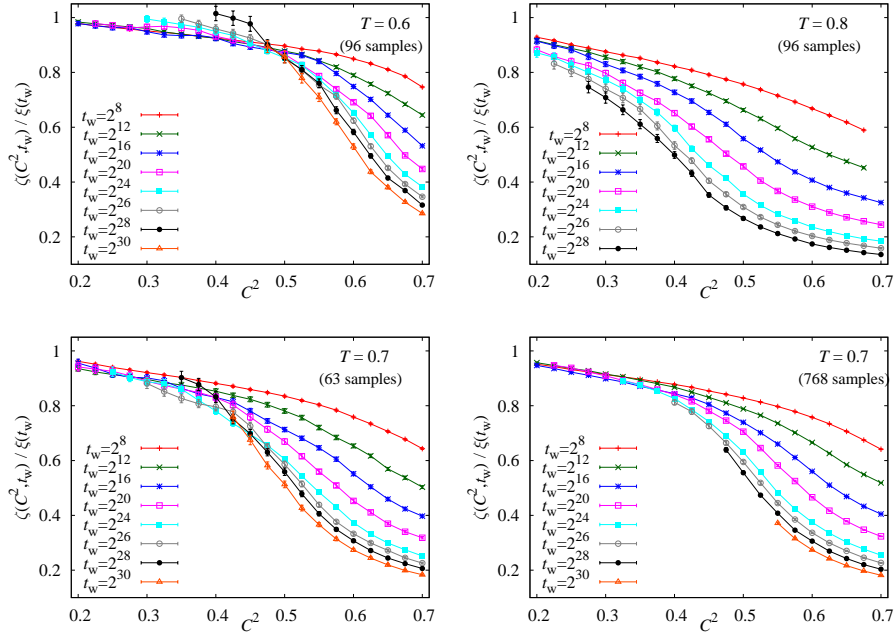}
\caption{Behavior of the ratio $R(C^2,\tw)=\zeta(C^2,\tw)/\xi(\tw)$ as a  function of $C^2$
  for several values of $\tw$. {\bf Top left:} at our lowest temperature
  $T=0.6$. {\bf Top right:} at temperature $T=0.8$.  {\bf Bottom left:} at
  temperature $T=0.7$ averaged over $63$ samples. {\bf Bottom right:} same plot
  as bottom left, averaging over $768$ samples.}
\label{fig:xiC2tw.over.xitw}
\end{center}
\end{figure}

In the large correlation sector, i.e.  $q_\text{EA}^2 < C^2$, and for large $\tw$,
the correlation length $\zeta(C^2,\tw)$ approaches a $\tw$ independent value.
On the other hand, one expects that for small values of $C^2$ (that is, $C^2 <
q_\text{EA}$) and large $\tw$, $\zeta(C^2,\tw)$ diverges as the coherence length
$\xi(\tw)$ defined in Eq.~(\ref{C4-DECAY})~\cite{LET-NSU}. Such behavior is represented
in Fig.~\ref{fig:xiC2tw.vs.C2}, in which we plot $\zeta$ as a function of
$C^2$, at temperatures $T=0.7$ and $T=0.8$, for some values of $\tw$. It is
also interesting to consider the ratio $R(C^2,\tw)=\zeta(C^2,\tw) / \xi(\tw)$
and study how its behavior as a function of $C^2$ changes with $\tw$. As
pointed out above, for small values of $C^2$ and large $\tw$, we expect
$R(C^2,\tw)\sim \text{const} > 0$. Moreover, since the coherence length $\xi(\tw)$
diverges for large $\tw$, $R(C^2,\tw)$ should vanish at large waiting times
when $C^2>q_\text{EA}^2$. In Fig.~\ref{fig:xiC2tw.over.xitw} we show $R(C^2,\tw)$ at
temperatures $T=0.6$, $0.7$ and $0.8$.  An interesting feature is the
crossover between the two sectors $C^2 < q_\text{EA}^2$ and $C^2 > q_{EA}^2$. At
$T=0.8$ the $q_\text{EA}$ is too small (see Table~\ref{TAB-QEA} in
Sect.~\ref{STATIONARY-C-SUBSECT} below) to let us observe the small $C^2$
behavior described above. On the other side, data for $R(C^2,\tw)$ at $T=0.6$
quickly approaches a constant value for small correlations. It seems that the
larger $\tw$, the fastest the convergence to a constant, determining in this
way a crossing point. However, these data suffer from large fluctuations that
do not allow us to make any strong speculation on the crossings among curves
at large values of $\tw$. Indeed, we know that $R(C^2,\tw)$ should vanish for
large $C^2$ roughly as $1/\xi(\tw)$, but up to our knowledge, there is no
reason for $R$, as a function of $C^2$, to converge faster to a constant when
$\tw$ increases.  In the bottom pictures of Fig.~\ref{fig:xiC2tw.over.xitw} we
report the same data at $T=0.7$, averaged on both $63$ (left) and $768$
samples (right). Even if simulations on larger sample statistics are not as
long as those of the first $63$ samples, the smoothing in the curves does not
improve the crossing definition. In addition, we have not been able to find
any clear scaling behavior of the crossing points with the waiting time.

Since at large times and small correlations $\zeta(C^2,\tw)$ and $\xi(\tw)$
only differ in a constant factor (which is also close to $1$), one may expect
that the behavior of the two-time, two-site correlators (Eq.~(\ref{C22-DEF})) should
be analogous to that of the four point correlation function (Eq.~(\ref{C4-DEF})).
We can then probe the long distance scaling of $C_{2+2}$ with the same
analysis performed in reference~\cite{DYNJANUS} (and in Sect.~\ref{SECT-a})
for $C_4(r,\tw)$, Eq.~\eqref{C4-DECAY}. In particular, we can extract the
exponent $b$ of the algebraic prefactor in Eq.~\eqref{C22-DECAY} and the dynamic
exponent $z_\zeta$, assuming a power-law growth for the correlation length:
\begin{equation}
  \zeta(C^2,\tw)=A\ \tw^{1/z_\zeta}\ .
\end{equation}
We fix $C^2$ to a $T$-dependent value, which is a value small enough to
be below the $q_\text{EA}^2$ at all considered temperatures $T=0.6$,
$0.7$ and $0.8$ ($C^2$ values should be compared with $q_{EA}^2$
estimates given in Sect.~\ref{STATIONARY-C-SUBSECT}), and large enough
to have the necessary number of $\tw$ points in order to obtain fair
fits. We also had to impose a $\zeta_{1,2} \geq 3$ constraint to avoid
the effects of lattice discretization. We did not impose any upper
limit to $\zeta$, whose values hardly reach $8$ lattice spacings at
$T=0.8$ (even less at colder temperatures), and we expect that in the
time sector considered the constraint imposed on $\xi$ in
reference~\cite{DYNJANUS} and in Sect.~\ref{SECT-a} would work as well
as in that case.

\begin{table}
\centering
\begin{tabular*}{\columnwidth}{@{\extracolsep{\fill}}ccrlcc}
\hline
$T$ &
$C^2$ &
\multicolumn{1}{c}{ $z_\zeta$} &
\multicolumn{1}{c}{ $b$} & $\chi^2_{\zeta}$/d.o.f. & $\chi^2_{I_1}$/d.o.f.\\
\hline
\multirow{2}{0.5cm}{$0.6$}
 & $0.200$ & $13.4(6)$ & $0.43(4)$ & $0.01/2 $ & $0.13/2$\\
 & $0.325$ & $12.8(4)$ & $0.55(3)$ & $7.16/7 $ & $4.45/7$\\
\hline
\multirow{2}{0.5cm}{$0.7$}
 & $0.200$ & $11.14(20)$ & $0.508(17)$ & $0.69/3 $ & $0.37/3$\\
 & $0.325$ & $11.35(12)$ & $0.642(9)$  & $8.08/7 $ & $7.70/7$\\
\hline
\multirow{2}{0.5cm}{$0.8$}
 & $0.100$ & $9.56(17)$  & $0.497(13)$ & $3.73/5$  & $3.37/5$\\
 & $0.175$ & $10.12(13)$ & $0.540(10)$ & $7.15/8$  & $7.91/8$\\
\hline
\end{tabular*}
\caption{Value of the dynamic exponent $z_\zeta$ and exponent $b$ for the algebraic prefactor, for
  three subcritical temperatures. We limited the fitting window
  constraining the fits to $\zeta_{1,2} \geq 3$. Data at $T=0.7$ averaged
  over $768$ samples.}
\label{TAB-b}
\end{table}

We summarize the obtained values of $b$ in Table~\ref{TAB-b}, reporting
results for the smallest $C^2$ attainable (that is, permitting
reasonable fits) at all temperatures, as well as for larger values of
$C^2$, for which the number of $t_w$ points allows for quite good fits.
We see that $b$ and $z_\zeta$ are slightly different from the values of $a$ and $z$
for the four point correlators presented in Sect.~\ref{SECT-a}.
Unfortunately our data do not permit a more precise determination for
these exponents in the deep $C^2 < q_\text{EA}^2$ sector. In this
respect, the determination of $q_{EA}$ is a crucial issue (see
Sect.~\ref{STATIONARY-C-SUBSECT}).

\section{The time correlation functions}\label{TIME-CORRELATION-FUNCTIONS-SECT}

\subsection{The stationary part of $C(t,\tw)$}\label{STATIONARY-C-SUBSECT}
\begin{figure}[t]
\centering
\includegraphics[height=.8\textwidth,angle=270]{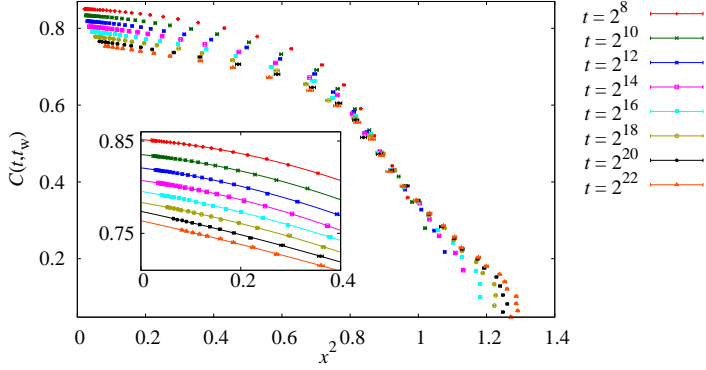}
\caption{The spin-spin correlation $C(t,\tw)$ as a function of
$x^2=\bigl(\zeta(t,\tw)/\xi(\tw)\bigr)^2$ for $T=0.7$. \textbf{Inset:}
Close up of the small $x^2$ region and comparison with our
quadratic fits.}\label{FIG-X2}
\end{figure}

\begin{figure}
\centering
\includegraphics[height=\textwidth,angle=270]{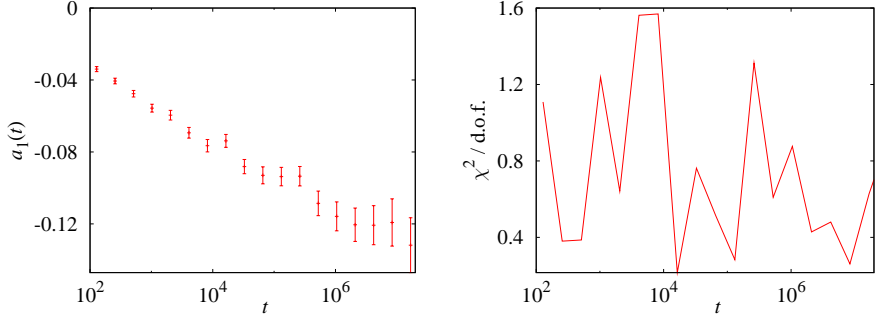}
\caption{\textbf{Left:} Slope at $x^2=0$ of the fitting
curves~\eqref{FIT-X2} at $T=0.7$ as a function of time.
\textbf{Right:} Check of the fit quality using
a $\chi^2$ estimator that takes into account correlations
only at equal $(t,\tw)$, but disregards correlations
among different pairs of times.}\label{FIG-X2-JI2}
\end{figure}

The naive computation of the stationary part of $C(t,\tw)$, $C_\infty(t)$ Eq.~\eqref{Cinf},
suffers from an essential problem: how to know when $\tw$ is large enough.
The consideration of characteristic length scales may simplify this problem,
as the limit $\tw \to\infty$ is equivalent to $\xi(\tw)^{-1}\to0$.\footnote{%
In what follows we shall always use the $\xi_{1,2}$ estimator.}
However, the approach to this limit will be acutely $t$-dependent.
Hence, it is better to consider a dimensionless variable,
\begin{equation}\label{EQ-X}
x(t,\tw) = \frac{\zeta(t,\tw)}{\xi(\tw)}\ .
\end{equation}
As we saw in Sect.~\ref{XIC2TW-SECT}, the correlation
length $\zeta(t,\tw)$ quickly reaches a $\tw$-independent
limit, so  $x(t,\tw)$ is essentially $\xi^{-1}(\tw)$
in its natural units for each $t$. In fact,
see Fig.~\ref{FIG-X2} for our data at $T=0.7$,
the plot of $C(t,\tw)$ against $x^2(t,\tw)$
is pretty smooth for $x^2\to0$. Furthermore,
the curves for different $t$ become parallel
as $t$ grows, which suggests the existence
of a smooth scaling function,
$C(t,\tw) = C_\infty(t) + f(x^2)$.
We have fitted the curves $C(t,x^2)$
for each $t$ in the range $x^2 \leq 0.5$
to a quadratic polynomial (see inset to Fig.~\ref{FIG-X2}),
\begin{equation}\label{FIT-X2}
C(t,x^2) = C_\infty(t) + a_1(t) x^2 + a_2(t) x^4.
\end{equation}
Unlike our treatment of the exponent $a$ (Sect.~\ref{SECT-a}), here
we cannot skirt the issue of errors in both coordinates.
As the effect of both errors is similar, we have performed
a least squares fit and an a posteriori $\chi^2$ test, Fig.~\ref{FIG-X2-JI2}, right.
This $\chi^2$ test was diagonal in the sense that we did not consider
correlations among different pairs of times, but we did consider
the covariance matrix for $(x^2,C)$ at the same $(t,\tw)$.
Our statement that the scaling curves in Fig.~\ref{FIG-X2} become parallel is made
quantitative in Fig.~\ref{FIG-X2-JI2}, left, which plots
the coefficient $a_1(t)$ in equation~\eqref{FIT-X2}.

\begin{figure}
\centering
\includegraphics[height=.7\textwidth,angle=270]{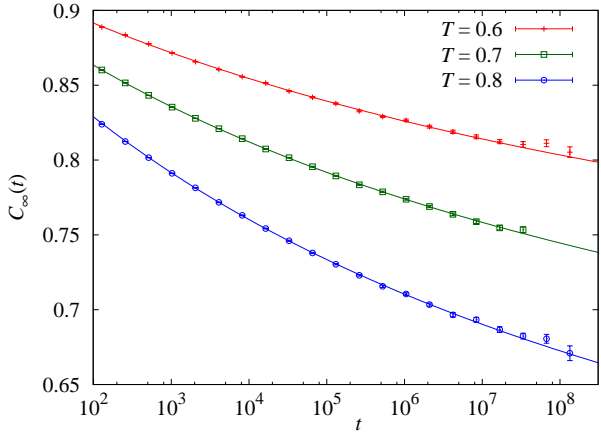}
\caption{Stationary component of $C(t,\tw)$, $C_\infty(t)$,
obtained with the extrapolations in Figs.~\ref{FIG-X2} and~\ref{FIG-X2-JI2},
for all our subcritical temperatures.}\label{Ct}
\end{figure}

The method described above has allowed us to compute $C_\infty(t)$
with remarkable accuracy for $t\lesssim10^8$, see Fig.~\ref{Ct}.
We can now try to perform a second extrapolation,
Eq.~\eqref{EQ-QEA}, and obtain the value of $q_\text{EA}$
for each of our simulated temperatures. In order to do this
we first tried a power law extrapolation
\begin{equation}\label{POWER-LAW}
C_\infty(t) = q_\text{EA} + A t^{B}.
\end{equation}
This functional form yielded very good fits for $T=0.6,0.8$
but the values of the exponent $B$ where very small, of about $B\sim -0.05$.
Unfortunately, the smallness of $B$ makes the extrapolation extremely
risky. Just to be on the safe side and check explicitly that $q_\text{EA}>0$,
we have tried a logarithmic fitting function,
\begin{equation}\label{LOGARITHM}
C_\infty(t) = q_\text{EA} + \frac{A}{B + \log t}\ .
\end{equation}
This Ansatz lacks any theoretical basis, but since it is
slower than any power law, we expect it to provide
a lower bound on $q_\text{EA}$. On the numerical side,
the logarithmic fit was as good as the power law (as determined by a $\chi^2$ test).
Nevertheless, as expected, it produced values of $q_\text{EA}$ which were incompatible
with those of equation~\eqref{POWER-LAW} (see Table~\ref{TAB-QEA}).
Furthermore, when we tried both extrapolating methods
for $T=0.7$, where we had simulated many more samples, we found
that they were somewhat forced. This leads us to conclude that
the real asymptotic behavior of $C_\infty(t)$ is probably something
in between equations~\eqref{POWER-LAW} and~\eqref{LOGARITHM}.
We shall use the difference between both methods
with a fitting window of $t\in[10^3,10^8]$ as our uncertainty interval,
\begin{align}
0.62&\leq q_\text{EA}(T=0.6) \leq 0.733,\nonumber \\
0.474&\leq q_\text{EA}(T=0.7)\leq 0.637, \label{COTAS-QEA}\\
0.368&\leq q_\text{EA}(T=0.8) \leq 0.556.\nonumber
\end{align}
Even with our unprecedentedly long simulations, we are still at the threshold
of being able to compute $q_\text{EA}$.

\begin{table}
\centering
\begin{tabular*}{\columnwidth}{@{\extracolsep{\fill}}cclcllc}
\hline
\multirow{2}{0.5cm}{$T$} &
\multirow{2}{1cm}{Fitting range}&
\multicolumn{2}{c}{ Logarithm} &
\multicolumn{3}{c}{ Power law} \\
& &
\multicolumn{1}{c}{$q_\text{EA}$} & $\chi^2/\text{d.o.f}$ &
 \multicolumn{1}{c}{$q_\text{EA}$} &\multicolumn{1}{c}{$-B\times10^2$} &  $\chi^2/\text{d.o.f}$ \\ 
\hline
\multirow{3}{0.5cm}{$0.6$} &
  $[10^2,10^8]$   & 0.607(16) & 34.1/17 & 0.730(8)  & 5.7(4)  &  31.2/17 \\ 
& $[10^3,10^8]$   & 0.62(3)   & 7.23/14 & 0.733(14) & 5.8(7)  &  7.59/14 \\
& $[10^4,10^8]$   & 0.62(5)   & 6.25/10 & 0.726(24) & 5.4(12) &  6.32/10 \\
\hline                                                           
\multirow{3}{0.5cm}{$0.7$} &                                     
  $[10^2,10^8]$   & 0.497(10) & 23.7/17 & 0.656(5)  & 6.16(18) &  32.6/17 \\
& $[10^3,10^8]$   & 0.474(21) & 18.9/14 & 0.637(11) & 5.5(3)   &  18.5/14 \\
& $[10^4,10^8]$   & 0.49(5)   & 15.0/10 & 0.63(3)   & 5.4(9)   &  15.3/10 \\
\hline                                                           
\multirow{3}{0.5cm}{$0.8$} &                                     
  $[10^2,10^8]$   & 0.371(13) & 6.50/17 & 0.568(7)  & 6.56(20)  &  9.39/17 \\
& $[10^3,10^8]$   & 0.368(24) & 5.53/14 & 0.556(12) & 6.2(4)    &  4.27/14 \\
& $[10^4,10^8]$   & 0.40(6)   & 4.31/10 & 0.56(3)   & 6.4(11)   &  3.82/10 \\
\hline
\end{tabular*}
\caption{Estimate of $q_\text{EA}$ for three subcritical temperatures,
using two different extrapolating functions. For $T=0.6,0.8$ both
are very good, but at $T=0.7$ (where we have better statistics)
they are somewhat forced. This suggests that the real $q_\text{EA}$ probably
lies in between our two estimates. For the power law
extrapolation, Eq.~\eqref{POWER-LAW}, we also quote the exponent $B$.
Notice that this exponent is not proportional to $T$.
}\label{TAB-QEA}
\end{table}

\subsection{Energy density and scaling exponents}\label{SECT-ENERGY}
The time decay of the energy density offers further insights into the
connection between statics and dynamics (see Sect.~\ref{XIGROWTH}).
At all temperatures the $L=80$ energy data are well described by a power-law
decay
\begin{equation}
\label{eq:e_decay}
e(\tw)-e_{\infty}=A\tw^{-\epsilon(T)}\ \ ,
\end{equation}
with $e_{\infty}$ the asymptotic (equilibrium) energy value.
At the critical temperature $T_{\mathrm c}$, general scaling arguments relate the
decay exponent $\epsilon$ to the  dimension of the energy operator $d_e=(D\nu -
1)/\nu$ and
the critical dynamic exponent $z$~\cite{PRRR}\ :
\begin{equation}
\label{eq:e_decay2}
e(\tw;T_\text{c})-e_{\infty;T_{\mathrm c}}=A\,\tw^{-d_e/z_\text{c}}\ \ .
\end{equation}
We could then in principle extract $\epsilon(T_{\mathrm c}=1.1)$ and compare with
the best estimate available (considering $\nu=2.45$ by Hasenbusch
\emph{et al.\/}~\cite{PELISSETTO} and $z_\text{c}=z(T_{\mathrm c})=6.86$~\cite{DYNJANUS}
we expect $\epsilon(T_{\mathrm c})\simeq0.378$).
Unfortunately we do not have enough statistics at the critical
temperature to allow for fair fits and statistical accuracy in the
determination of the decay exponent, but we
can avoid such difficulty.
Assuming critical dynamics in all the Spin Glass phase ($T < T_{\mathrm c}$)
such relation could be extended and tested at all simulated
temperatures, provided a $T$-dependent
dynamic exponent $z(T)$ is considered (see Sect.~\ref{XI-SECT} above). In addition, given
an inversely proportional dependence of $z(T)$ on the temperature, we
expect that $\epsilon(T) \propto T$. Once the exponent has been
determined for some values of $T<T_{\mathrm c}$, we can extrapolate to determine $\epsilon(T_{\mathrm c}^-)$.
The exponent may be in principle
discontinuous at the critical point, so some analysis has been
performed also at $T=1.15$, slightly higher than $T_{\mathrm c}$, to have a hint
on the possible value of the right limit $\epsilon(T_{\mathrm c}^+)$.
\begin{figure}[b]
\begin{center}
\includegraphics[height=0.65\textwidth,angle=270]{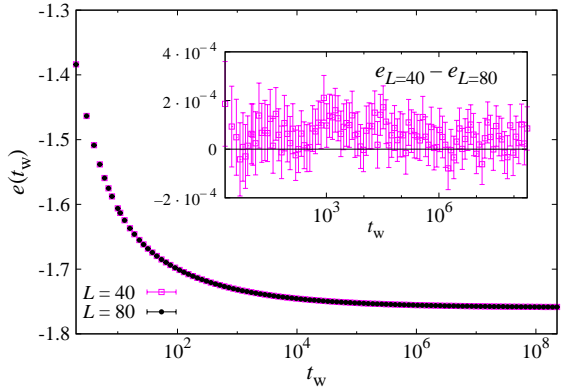}
\caption{No appreciable dependence of the decay $e(\tw)$ on the system size,
  for lattices $L=40$ and $L=80$ at $T=0.8$. \textbf{Inset:} the difference between the two
  as a function of Monte Carlo time.}
\label{fig:e40e80}
\end{center}
\end{figure}

At all temperatures a fit to power law Eq.~(\ref{eq:e_decay}) is quite sensitive
to the fitting window. This is not surprising as the early dynamics may be
very different from the asymptotic behavior. In addition, we are aware that
our data may suffer from important finite size effects when the coherence
length grows up over some fraction of the system size (see~\cite{DYNJANUS}).
The limit in $\tw$ beyond which data for some observable cannot be
considered in the thermodynamic limit depends on temperature and on both
the observable and the precision with which it can be measured (which for the energy is
very high, of order one part in $10^5$).  At each temperature, the upper limit in
the fitting window should then in principle depend on when, in Monte Carlo
time, we start experiencing finite size effects in the energy. We can check
this by comparing with the energy decay of systems of size $L=40$. As an
example, in Fig.~\ref{fig:e40e80} we report the difference
$e_{L=40}(\tw)-e_{L=80}(\tw)$ at $T=0.8$ as function of $\tw$: no appreciable deviations
appear in the whole simulated range, allowing us to extend our fit to the
largest available time.  We are then left with adjusting the lower limit
$t_\text{min}$ by checking that the fit parameters $e_{\infty}$, $A$ and $\epsilon$
go to stable values.

\begin{figure}[t]
\begin{center}
\includegraphics[height=\textwidth,angle=270]{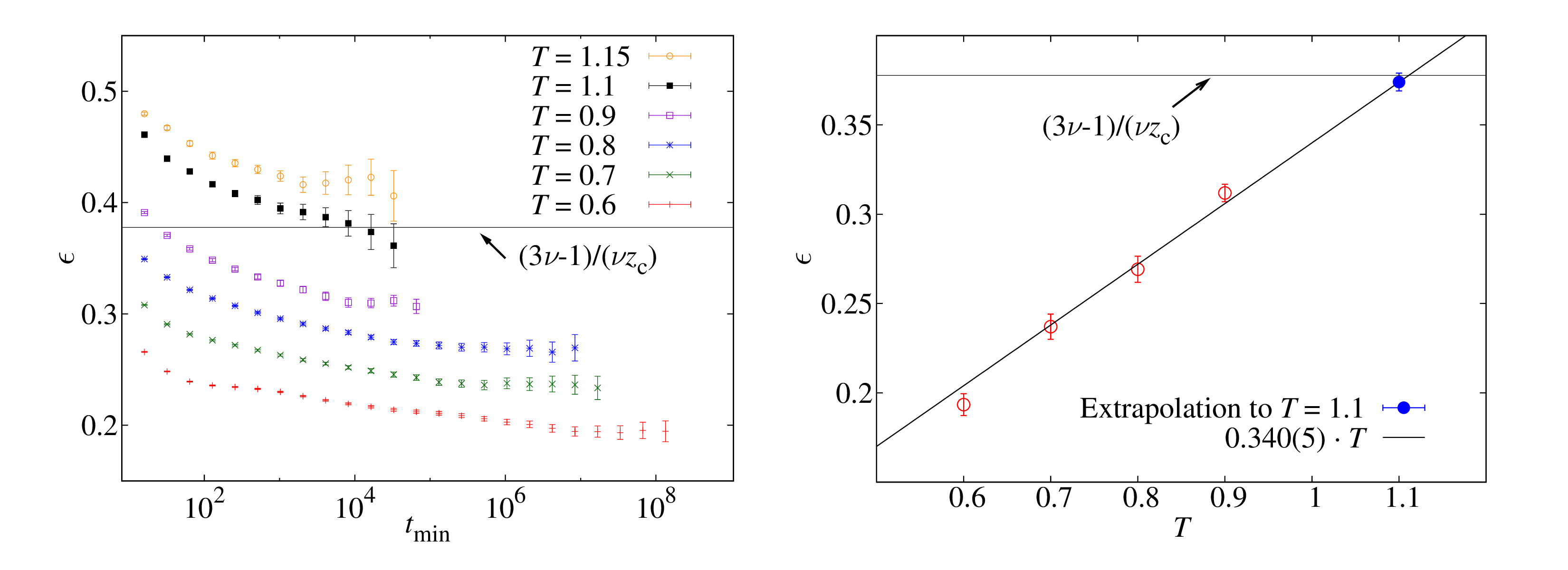}
\caption{\textbf{Left:} dependence of $\epsilon$ on the fitting window. For
  each choice of the lower limit $t_\text{min}$ we fit for all points $\tw >
  t_\text{min}$. We do not assume any upper limit in the fitting window
  (see text). At lower temperatures we find $t_\text{min}$-independent
  parameters, with clear plateaus showing up. For each curve the plateau value is
  taken considering the midpoint in the last longer series of an even
  number of points lying on an horizontal line within error bars.
  At $T\geq 1.1$ no clear plateau appears, while large fluctuations
  spoil the fits. The horizontal line is a prediction for
  $\epsilon(T_{\mathrm c})$ considering the value of the critical exponent $\nu$
  of~\cite{PELISSETTO} and $z_\text{c}=z(T_{\mathrm c})=6.86$ from Table~\ref{TAB-a}.
  \textbf{Right:} the plateau values for $\epsilon(T)$ for simulated values of
  $T\leq 0.9$ as function of $T$ (empty circles, red color online). A proportionality law $\epsilon(T) =
  cT$ works well, and allows one to extrapolate $\epsilon$ up to
  $T_{\mathrm c}=1.1$ (the upper full circle, blue color online). The horizontal line
  is the best estimate by taking the $\nu$ value of~\cite{PELISSETTO}
  and $z_\text{c}$ from Table~\ref{TAB-a}.}
\label{fig:e_decay}
\end{center}
\end{figure}

\begin{table}
\centering
\begin{tabular*}{\columnwidth}{@{\extracolsep{\fill}}clcc}
\hline
$T$ & \multicolumn{1}{c}{$E_{\infty}$ }& $A$ & $\epsilon$ \\
\hline
$0.6$ & $-1.778\,62(10)$ & $0.122(11)$ & $0.193(6)$ \\
$0.7$ & $-1.770\,84(7)\ \ \ \ \, $   & $0.148(15)$ & $0.236(7)$ \\
$0.8$ & $-1.759\,47(7)$ & $0.155(16)$ & $0.268(7)$ \\
$0.9$ & $-1.744\,29(13)\ \  $  & $0.19(1)\ \ \ \ $   & $0.312(5)$ \\
\hline
\end{tabular*}
\caption{Results for the power-law decay parameters, Eq.~(\ref{eq:e_decay}).}
\label{tab:e_decay}
\end{table}
Results for $\epsilon$ as function of $t_\text{min}$ at all
simulated temperatures are depicted in Fig.~\ref{fig:e_decay}, left. At low
temperatures it has been possible to probe for $t_\text{min}$ ranging in several
decades. At higher $T$, it is not possible with our data to have fair fits at
larger $t_\text{min}$ values, as very large fluctuations arise, thus hiding any
possible plateau (error bars are from a jackknife procedure).

At $T=0.6$, $0.7$, $0.8$ we considered the parameters converged in the last
five points in Fig.~\ref{fig:e_decay} left; at $T=0.9$ the last three points
converged within error bars. In any of these intervals, we take the mid-point
as the plateau values, and report them in the plot of Fig.~\ref{fig:e_decay}
right (corresponding values of all parameters are reported in
Table~\ref{tab:e_decay}).  The data are well represented by a
temperature-proportional law, with $\chi^2/\mbox{d.o.f.}=4.8/3$. The
extrapolated value of $\epsilon(T_{\mathrm c}^-)$ is $0.373(5)$ that coincides within
errors with the best estimate reported above. This is a quite good \emph{a
  posteriori} confirmation that the procedure described is robust, leastwise,
within the precision attainable.  From Fig.~\ref{fig:e_decay} left, we see
that at $T=1.1$, $1.15$ we were not able to identify a clear plateau and
estimate $\epsilon$ directly, still there is a trend towards the expected
critical value, supporting that $\epsilon(T)$ should be linear up to the
critical point at least.

From the values of $\epsilon(T)=(3\nu-1)/\bigl(\nu z(T)\bigr)$,\footnote{%
The formula is expected to be valid if the value of the exponents below the
critical temperature are controlled by their value at $T_\text{c}$.}
we obtain $z(0.6)=13.4(3)$, $z(0.7)=11.0(3)$ and
$z(0.8)=9.7(3)$ in agreement with the results of Table~\ref{TAB-a}.

Alternatively we can link these numerical results with an analytical one obtained by
Franz {\em et al.\/}~\cite{FRANZ2} (see also~\cite{PRL_MPRR}). In this reference, the contribution of the
interface was computed in the framework of Replica Symmetry
Breaking. Assuming the power law behavior of the coherence length, one would
expect that $\epsilon(T)=2.5/z(T)$. So, inputting our values for $z(T)$
we obtain $\epsilon(T)=0.33 T$, which is compatible with our numerical
finding ($\epsilon(T)=0.340(5) T$, see Fig.~\ref{fig:e_decay} (left)).

We conclude this section by testing in nonequilibrium simulations (see also~\cite{OUT}), the 
correction-to-scaling exponent measured by Hasenbusch {\em et al.\/}~\cite{PELISSETTO}
$\omega=1.0$. Our data at $T_{\mathrm c}=1.1$ are well described by a double
power-law (for $\tw>2^6$, $\omega=1.0$, $\epsilon=0.378$)
\begin{equation}
e(\tw)-e_{\infty}
= A\tw^{-\epsilon}\left(1+B\tw^{-\omega/z_{\text{c}}}\right),
\end{equation}
giving $e_{\infty}=1.70201(2)$, $A=0.149(2)$, $B=1.25(5)$ and
$\chi^2/\mbox{d.o.f} = 104.20/102$.

\subsection{The thermoremanent magnetization}
The thermoremanent magnetization of a SG has
been known since the 1980s to decay with a power
law~\cite{THERMOREMANENT1,THERMOREMANENT2}
(deviations to this simple behavior
were observed only extremely close to
$T_\text{c}$, namely $T>0.98T_\text{c}$).
As we said in Sect.~\ref{SECT-OBSERVABLES}, this observable
can be identified with $C(t,\tw)$ for fixed  $\tw$
and $t\gg \tw$. Following~\cite{THERMOREMANENT3},
 we have fitted our data
to a decay law
\begin{equation}\label{C-DECAY}
C(t,\tw) = A'(\tw) + B(\tw) t^{c(\tw)}\ .
\end{equation}
The constant term $A'(\tw)$ is justified
because for a finite number of samples,
the correlation function does not
go to zero for large times. The very same
problem arises in the analysis of experimental
data~\cite{THERMOREMANENT1} (it was solved by taking a numerical derivative).

We summarize the results of these fits
on Table~\ref{TAB-THERMOREMANENT}. We
fixed $\tw=2,4,8,16$ and fitted our data
up to the point where finite size
effects appear ($t< 2\times 10^{10}$ for $T=0.7$,
$t<4\times 10^8$ for $T=0.8$ and the whole range
for $T=0.6$). We started our fits at $t=10^6$,
but this limit can be varied along several
orders of magnitude with no change in the fitting
parameters or the goodness of the fit.
As we can see, the exponent $c$ has a slight, but systematic,
dependence on $\tw$. This tendency was already observed
by~\cite{KISKER}.

\begin{table}[b]
\centering
\begin{tabular*}{\columnwidth}{@{\extracolsep{\fill}}ccclllc}
\hline
$T$ &
$\tw$& $t_\text{min}$ &
\multicolumn{1}{c}{ $c(\tw)$} &
\multicolumn{1}{c}{ $d(\tw)$} &
\multicolumn{1}{c}{ $A'(\tw)\times10^3$}&
  $\chi^2$/d.o.f.\\
\hline
\multirow{4}{0.5cm}{$0.6$}
&2 &  $10^6$ &  $-0.1525(23)$ & $2.14(5)$ & 2.6(6)  & 14.7/64\\
&4 &  $10^6$ &  $-0.1495(22)$ & $2.10(5)$ & 2.8(8)  & 15.5/64\\
&8 &  $10^6$ &  $-0.1459(20)$ & $2.05(4)$ & 2.5(10) & 17.4/64\\
&16&  $10^6$ &  $-0.1430(19)$ & $2.01(4)$ & 2.4(12) & 17.5/64\\
\hline
\multirow{4}{0.5cm}{$0.7$}
&2 &  $10^6$ &  $-0.1787(14) $ & $2.067(27)$ & 1.47(25) & 23.3/50\\
&4 &  $10^6$ &  $-0.1765(13) $ & $2.041(26)$ & 1.8(3)   & 18.4/50\\
&8 &  $10^6$ &  $-0.1733(12) $ & $2.004(25)$ & 1.7(4)   & 18.9/50\\
&16&  $10^6$ &  $-0.1704(12) $ & $1.971(25)$ & 1.6(5)   & 15.4/50\\
\hline
\multirow{4}{0.5cm}{$0.8$}
&2 &  $10^6$ &  $-0.210(8)$    & $1.98(9)$   & 1.7(10)  & 13.9/32\\
&4 &  $10^6$ &  $-0.212(7)$    & $2.00(8)$   & 2.8(12)  & 11.1/32\\
&8 &  $10^6$ &  $-0.208(7)$    & $1.96(8)$   & 3.0(14)  & 10.8/32\\
&16&  $10^6$ &  $-0.205(6)$    & $1.93(7)$   & 3.0(18)  & 8.43/32\\
\hline
\end{tabular*}
\caption{Exponents $c$ and $d$ and parameter $A'$ of equations~\eqref{C-DECAY}
and~\eqref{C-XI}. The values of $c$ come from
fits, whose $\chi^2$ we also show. The exponent
$d$ comes from $d=-cz$, where $z$ is the dynamic
exponent of equation~\eqref{XI-GROWTH} and Table~\ref{TAB-a}.
Fits are limited to the time
range where the system is free of finite size effects,
which accounts for the different numbers of degrees of freedom
for each temperature.}\label{TAB-THERMOREMANENT}
\end{table}

Our estimates for $c$ can be compared with experimental
values~\cite{THERMOREMANENT1}
\begin{align}
c( T=0.55T_\text{c} ) &\approx -0.12,\nonumber\\
c( T=0.67T_\text{c} ) &\approx -0.14,\label{c-EXP}\\
c( T=0.78T_\text{c} ) &\approx -0.17,\nonumber
\end{align}
(the error bars on these exponents are small,
 not much larger than the size of the plotted data points in Fig. 3b
of~\cite{THERMOREMANENT1}).
When compared with the values in Table~\ref{TAB-THERMOREMANENT},
the experimental exponents are similar but slightly higher
(let us recall that $0.6 \approx 0.55 T_\text{c}$,
$0.7\approx 0.64T_\text{c}$ and $0.8\approx 0.73T_\text{c}$).

A different approach comes from realizing that in experimental
work $t$ and $\tw$ typically differ by at most
4 orders of magnitude, while in our fits they differ by
as many as $9$ or $10$. Taking this into account, it is interesting
to consider a power law where the fitting window is
shifted with $\tw$ so that $1\leq t/\tw\leq 10$~\cite{DYNJANUS},
\begin{equation}\label{FULL-AGING}
C(t,\tw) = A(\tw) \left(1 + t/\tw\right)^{-1/\alpha(\tw)}.
\end{equation}
Using this functional form and extrapolating
$-1/\alpha(\tw)$ to a typical experimental time
with a quadratic fit we
obtain
\begin{align}
-1/\alpha(\tw=100\text{ s}) &\approx -0.11,& T &= 0.6 \approx 0.55T_\text{c},\nonumber\\
-1/\alpha(\tw=100\text{ s}) &\approx -0.12,& T &= 0.7 \approx 0.64T_\text{c},\label{ALPHA}\\
-1/\alpha(\tw=100\text{ s}) &\approx -0.14,& T &= 0.8 \approx 0.73T_\text{c}.\nonumber
\end{align}
These extrapolations are slightly above the experimental values of
equation~\eqref{c-EXP}, but if both sets of exponents are interpolated
with a parabola, the two curves result roughly parallel, i.e.,
our extrapolation error seems temperature-independent.

Since both the thermoremanent magnetization, Eq.~\eqref{C-DECAY},
and the coherence length are well described by a power law,
it follows that $C(t,\tw)$ should be a power of $\xi(t+\tw)$,
at least for the small values of $\tw$ in Table~\ref{TAB-THERMOREMANENT},
\begin{equation}\label{C-XI}
C(t,\tw) \sim \xi(t+\tw)^{-d}.
\end{equation}
Indeed, following the same procedure we used for the
exponent $a$ of equation~\eqref{C4-DECAY} (see Sect.~\ref{SECT-a})
we have computed the values of $d$ for $\tw=2,4,8,16$ (Table~\ref{TAB-THERMOREMANENT}),
obtaining $d\approx 2$. Incidentally, let us remark that
the coherence length is notoriously difficult to access experimentally~\cite{ORBACH},
while the thermoremanent magnetization is a pretty standard measurement. Eq.~\eqref{C-XI}
then appears as an interesting, albeit indirect, way of estimating experimentally
an effective coherence length.

\begin{figure}
\centering
\includegraphics[height=.65\linewidth,angle=270]{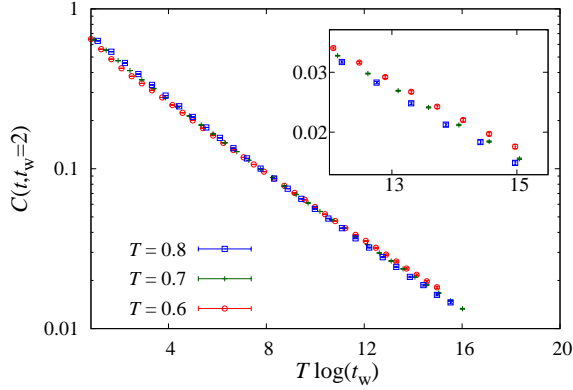}
\caption{$C(t,\tw=2)$ as a function of $T\log(t)$ for three
subcritical temperatures. The inset shows that the scaling
holds only approximately.}\label{REMANENTE-LOG}
\end{figure}

Notice that the exponents $c(\tw)$ of Table~\ref{TAB-THERMOREMANENT}
are roughly, but not exactly, linear in $T$. This suggests that
the thermoremanent magnetization should be a temperature-independent
function of $T \log(t)$~\cite{THERMOREMANENT2}. Fig.~\ref{REMANENTE-LOG}
shows that this is only an approximate claim.

\begin{table}
\centering
\begin{tabular*}{\columnwidth}{@{\extracolsep{\fill}}cccllc}
\hline
$T$ &
$\tw$&
$t_\text{min}$ &
\multicolumn{1}{c}{$e(\tw)$} &
\multicolumn{1}{c}{$f(\tw)$} &
$\chi^2$/d.o.f.\\
\hline
\multirow{8}{0.5cm}{$0.6$} 
&\multirow{2}{0.3cm}{$2$} &
      $10^3$ & $-0.236(7)$ & $0.873(9)$ & 52.2/104\\ 
& &   $10^6$ & $-0.30(6)$  & $0.82(5)$ & 13.9/64\\
&\multirow{2}{0.3cm}{$4$} &
      $10^3$ & $-0.203(6)$ & $0.909(8)$ & 47.9/104\\
 & &  $10^6$ & $-0.25(4)$  & $0.85(4)$ & 13.5/64\\
&\multirow{2}{0.3cm}{$8$} &
      $10^3$ & $-0.176(4)$ & $0.943(7)$ & 41.5/104\\
& &   $10^6$ & $-0.21(3)$  & $0.90(4)$ & 13.9/64\\
&\multirow{2}{0.3cm}{$16$} &
      $10^3$ & $-0.158(4)$ & $0.968(7)$ & 38.1/104\\
& &   $10^6$ & $-0.19(3)$ & $0.92(4)$ & 15.3/64\\
\hline
\multirow{8}{0.5cm}{$0.7$} 
&\multirow{2}{0.3cm}{$2$} &
      $10^3$ & $-0.263(4)$ & $0.890(4)$ & 43.0/90\\
& &   $10^6$ & $-0.32(3)$ & $0.84(3)$ & 14.4/50 \\
&\multirow{2}{0.3cm}{$4$} &
      $10^3$   & $-0.230(3)$ & $0.921(4)$ & 71.9/90\\
& &   $10^6$   & $-0.29(3)$  & $0.862(25)$ & 12.8/50\\
&\multirow{2}{0.3cm}{$8$} &
     $10^3$    & $-0.2003(23)$ & $0.955(3)$ & 94.1/90\\
& &  $10^6$    & $-0.253(23)$  & $0.895(23)$ & 13.0/50\\
&\multirow{2}{0.3cm}{$16$} &
      $10^3$   & $-0.1768(20)$ & $0.985(3)$ & 138/90\\
& &   $10^6$   & $-0.226(19)$  & $0.921(22)$ & 10.6/50\\
\hline
\multirow{8}{0.5cm}{$0.8$} 
&\multirow{2}{0.3cm}{$2$} &
      $10^3$   & $-0.302(16)$  & $0.891(16)$ & 45.1/72\\
& &   $10^6$   & $-0.5(3)$     & $0.77(15)$ & 14.3/32\\
&\multirow{2}{0.3cm}{$4$} &
      $10^3$   & $-0.257(12)$ & $0.934(14)$ & 63.6/72\\
& &   $10^6$   & $-0.6(4)$    & $0.71(17)$ & 11.4/32\\
&\multirow{2}{0.3cm}{$8$} &
      $10^3$   & $-0.223(10)$ & $0.970(13)$ & 69.8/72\\
& &   $10^6$   & $-0.49(24)$  & $0.76(12)$ & 11.1/32\\
&\multirow{2}{0.3cm}{$16$} &
      $10^3$   & $-0.192(8)$ & $1.008(12)$ & 65.9/72\\
& &   $10^6$   & $-0.40(19)$ & $0.81(12)$ & 8.49/32\\
\hline
\end{tabular*}
\caption{Parameters of a fit to Eq.~\eqref{C-DECAY2}, 
offering an alternative description of the 
thermoremanent magnetization. For each temperature, 
the maximum time included in the fit was such that 
$\xi(t_\text{max}) = 10$.}
\label{TAB-THERMOREMANENT2}
\end{table}

The incompatibility of the values in Eqs.~\eqref{c-EXP} and~\eqref{ALPHA},
together with the fact that we needed the constant term $A'(\tw)$
in~\eqref{C-DECAY}, suggests that there probably exists
a systematic error in the power law fits of Table~\eqref{TAB-THERMOREMANENT}.\footnote{%
Note, however, that the smallest value of $C(t,\tw=2)$ that 
we reach is $\sim0.013$, pretty large as compared with $A'(\tw)$.}
One may consider an alternative description,
\begin{equation}\label{C-DECAY2}
C(t,\tw) = A''(\tw) \exp\left[ e(\tw) (\log t)^{f(\tw)}\right]\ ,
\end{equation}
that would reproduce a power law if $f(\tw)=1$. Eq.~\eqref{C-DECAY2}
should not be confused with the stretched exponential, discarded
in reference~\cite{THERMOREMANENT1} for all $T$ but the closest
to $T_\text{c}$. We show our results for this fit
in Table~\ref{TAB-THERMOREMANENT2}. From the point of 
view of a $\chi^2$ test, the two descriptions, \eqref{C-DECAY} 
and \eqref{C-DECAY2}, are equally good. The fit
parameters are remarkably stable with variations of the fitting
window. As we see, the values of
$f(\tw)$ are very close to, but incompatible with, $1$ 
(at least for the lowest temperatures).

\begin{figure}
\centering
\includegraphics[height=\linewidth,angle=270]{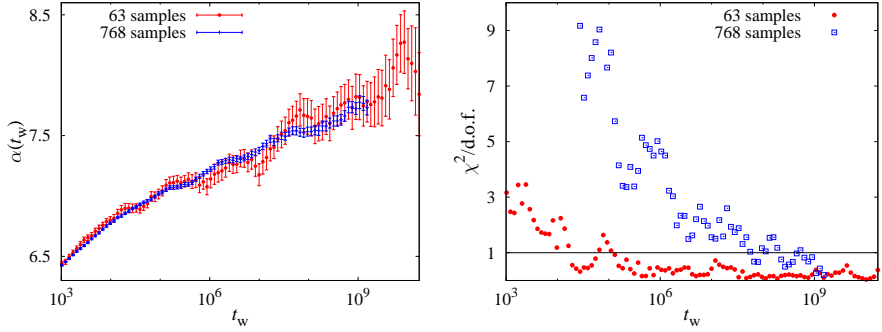}
\caption{Exponent $\alpha(\tw)$ defined in equation~\eqref{FULL-AGING}
as a function of $\tw$ at $T=0.7$, computed for the 63 samples of~\cite{DYNJANUS}
and for our 768 samples. The right panel shows the values
of diagonal $\chi^2$ for both fits.}
\label{FIG-WIGGLES}
\end{figure}

Let us conclude this section by considering again
the effects of statistical data correlation on
fit parameters. Specifically, let us consider
the exponent $\alpha(\tw)$ of equation~\eqref{FULL-AGING},
see Fig.~\ref{FIG-WIGGLES} and reference~\cite{DYNJANUS}.
The alert reader will notice strange wiggles, large as
compared with the error bars, which are specially prominent
for the fit with 63 samples. The reason is that
data for different $t$ and $\tw$ in this fit are
even more correlated than usual. In fact, we have sampled the function
$C(t,\tw)$ for $t$ and $\tw$ integer approximations to $2^{i/4}$, $i=0,1,\ldots$ For
each $\tw$, we perform the fit and extract $\alpha(\tw)$ for $\tw\leq t <10\tw$.
In other words, for each $\tw$ we used 14 values of $t$. Recall that
the spin configurations involved are the one at time $\tw$, and the 14
spin configurations at succeeding times $t+\tw$. Given our choice for
integer $t$ and $\tw$, it follows that, for the next $\tw$, in the
computation of $C$ we will use 13 out of the 14 spin configurations used
at the earliest $\tw$. For the second-next $\tw$, the number of repeated
configurations will be 12, and so forth. This is the origin of the
dramatic data correlation: the very same spin configurations are being
used for consecutive times. Notice that when the statistics
is increased to 768 samples, the period of these oscillations
does not change, but their amplitude decreases. Also in Fig.~\ref{FIG-WIGGLES},
right panel, we show the value of $\chi^2$/d.o.f. for the fits with
63 and 768 samples. In both cases, $\chi^2$ decreases strongly with $\tw$,
but, while the fit with 63 samples was perfectly reasonable from
$\tw\sim10^4$, that with 768 samples is not good until
$\tw \sim 10^8$. The increased accuracy reveals systematic deviations
from equation~\eqref{FULL-AGING}. Nevertheless, the estimate of $\alpha(\tw)$ for
both fits is compatible in a much wider range.

\subsection{$C_\text{link}$ as a function of $C^2$}
Studying $C_\text{link}$ as a function of $C^2$ for fixed $\tw$
we can monitor the scaling of the active domains' surface area.
This scaling has an exact correlate in equilibrium studies~\cite{ULTRA},
where one considers the probability distribution function of
\begin{equation}\label{QLINK}
Q_\text{link} = \frac{1}{ND} \sum_{\langle \vn x,\vn y\rangle}
 \sigma^{(1)}_{\vn x} \sigma^{(2)}_{\vn x}
 \sigma^{(1)}_{\vn y} \sigma^{(2)}_{\vn y}\ .
\end{equation}
The analogue of $C_\text{link}$ as a function of $C^2$
is the conditional expectation value $\langle Q_\text{link} | q^2\rangle$.
Indeed, in~\cite{DYNJANUS}, we quantitatively compared our dynamical results
with the equilibrium $\langle Q_\text{link} | q^2\rangle$ of~\cite{ULTRA}.
It was found that a convenient time-length dictionary
could be established in such a way that the equilibrium
results for finite lattices reproduced the nonequilibrium
results for finite $\xi(\tw)$. For instance, for $T=0.7$,
when $L/\xi(\tw)\approx 3.7$, $C_\text{link}(C^2,\tw)\simeq \langle Q_\text{link} | q^2\rangle_L$.

In~\cite{DYNJANUS} it was found that $C_\text{link}$ is a smooth
increasing function of $C^2$ at least up to experimental scales.
On the other hand, for systems undergoing coarsening dynamics (e.g.,
a disordered ferromagnet), $C_\text{link}$ tends to a constant $C^2$-independent
value whenever $C^2<q_\text{EA}^2$. Let us briefly
justify these expectations.

On the one hand, since in the RSB scenario the coherent domains
are not compact objects, one would expect $C_\text{link}$
to have the same aging properties as $C^2$, that is,
$\mathrm{d}C_\text{link}/\mathrm{d}C^2$ should not vanish. This is the nonequilibrium
analogue of the overlap equivalence property~\cite{ULTRA}.
For instance, in the Sherrington-Kirkpatrick model it is
straightforward to show that $C_\text{link}=C^2$.

On the other hand, to find the scaling for a coarsening image
of compact active droplets we need a more elaborate argument.
We consider a large droplet of size $\xi(t+\tw)$  at time $t+\tw$ that, at
time $\tw$, was made of $N_\text{C}$ smaller droplets of size $\xi(\tw)$.
The number of spins in the boundary of a droplet at time $\tw$
scales as $\xi(\tw)^{D_\text{s}}$. Typically, $D_\text{s}=D-1$,
but one may have $D-1\leq D_\text{s}\leq D$~\cite{SUPERUNIVERSALITY,DROPLET} (for
the TNT model of SG, $D-D_\text{s}\approx0.45$ for $D=3$, see~\cite{Ds} and Palassini and Young in~\cite{TNT}).
Of course, $N_\text{C}$ scales as $N_\text{C} \sim [\xi(t+\tw)/\xi(\tw)]^D$.
The overlap of each of the $N_\text{C}$ droplets at time $\tw$ with
the configuration at time $t+\tw$ is randomly $\pm q_\text{EA}$.
Hence, the scaling of $C(t,\tw)$ is, for the region $C<q_\text{EA}$,\footnote{%
  Even if Eq.~(\ref{C-COARSENING}) is
  intuitively evident, it can be backed by an explicit computation.
  From Eq.~(6-11) in~\cite{GI-JO-LE-94}, one easily shows for
  the {\em Ising} ferromagnet that the spin-spin correlation function takes
  the form of a series $C(t,\tw)=a_1 y+a_2 y^2+\ldots$, where
$y=\xi^{D/2}(t+\tw) \xi^{D/2}(\tw)/[\xi^2(t+\tw)+\xi^2(\tw)]^{D/2}\,.$
Eq.~(\ref{C-COARSENING}) follows when $\xi(t+\tw)\gg\xi(\tw)\,$.}
\begin{equation}\label{C-COARSENING}
C(t,\tw) \sim \sqrt{N_\text{C}} \left(\frac{ \xi(\tw)}{\xi(t+\tw)}\right)^{D}\sim
 \left(\frac{ \xi(\tw)}{\xi(t+\tw)}\right)^{D/2}\ .
\end{equation}
Now, for the link overlap we expect ($C_\text{link}^0$ is
the equilibrium expectation value of $Q_\text{link}$)
\begin{equation}
C_\text{link}(t,\tw)= C_\text{link}^0 + N_\text{C} \frac{\xi^{D_\text{s}}(\tw)}{\xi^D(t+\tw)}.
\end{equation}
In fact, the decay of $C_\text{link}(t,\tw)$ comes mainly from
the contribution of droplets' surface at time
$\tw$. In particular, for $t\to\infty$, the excess of $C_\text{link}$
over $C^0_\text{link}$ is just the probability that a link belongs
to the surface of a droplet at time $\tw$. Now, considering
equation~\eqref{C-COARSENING} we conclude that the number
of droplets $N_\text{C}$ scales with $C(t,\tw)$ as
\begin{equation}\label{NC-SCALING}
N_\text{C} \sim \frac{g(C)}{C^2}\, ,
\end{equation}
where the function $g(C)$ is continuous, but not necessarily differentiable at $C=0$,
and where $g(0) > 0$. Combining Eq.~\eqref{NC-SCALING} and Eq.~\eqref{C-COARSENING}
we get
\begin{equation}\label{CLINKC2-COARSENING}
C_\text{link}(t,\tw) = C_\text{link}^0 + C_\text{link}^1 g(C)  \xi^{D_\text{s}-D}(\tw)\ .
\end{equation}
In the above expression $C_\text{link}^1$ is a constant.
Notice that, in particular, equation~\eqref{CLINKC2-COARSENING} implies
that the derivative of $C_\text{link}$ with respect
to $C^2$ goes to zero as $\tw\to\infty$.

\begin{figure}
\centering
\includegraphics[height=\linewidth,angle=270]{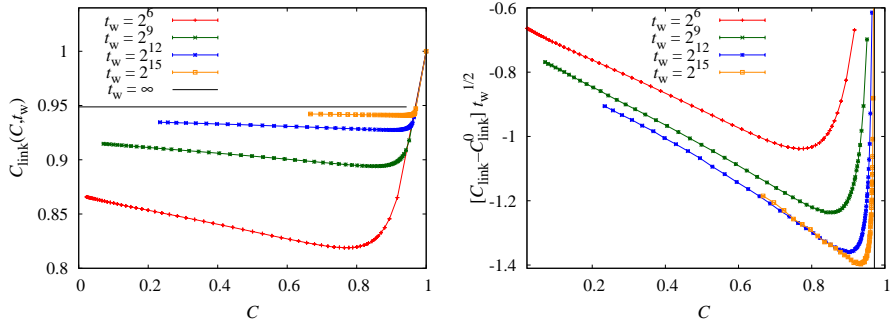}
\caption{\textbf{Left:} $C_\text{link}$ as a function of $C$
for the two-dimensional ferromagnetic Ising model at  $T=0.66T_\text{c}$.
Results obtained for an $L=4096$ lattice (the results were
averaged over $20$ trajectories). \textbf{Right:} Numerical
check of Eq.~\eqref{CLINKC2-COARSENING} for the data on the left panel.
The vertical line is at $C=q_\text{EA}$.
We see that, for large $\tw$ and $C<q_\text{EA}$, $C_\text{link}(C,\tw)-C_\text{link}^0$ scales
as $\xi^{-1}$.}
\label{CLINK-ISING}
\end{figure}

We can easily check Eq.~\eqref{CLINKC2-COARSENING} for the
two-dimensional Ising model, where the Onsager and Yang solutions
provide exact values for $C_\text{link}^0$ and $q_\text{EA}$
(in this case $D-D_\text{s}=1$). As for the growth of the coherence
length, it scales as $\tw^{1/2}$ (see, for instance~\cite{BRAY}).
As expected, see Fig.~\ref{CLINK-ISING}, $C_\text{link}$ tends
to a constant function $C_\text{link}(C,\tw\to\infty)=C_\text{link}^0$
for $C< q_\text{EA}$. As we show in Fig.~\ref{CLINK-ISING}, right,
the approach to this constant is well described by Eq.~\eqref{CLINKC2-COARSENING}.

\begin{figure}
\centering
\includegraphics[height=\linewidth,angle=270]{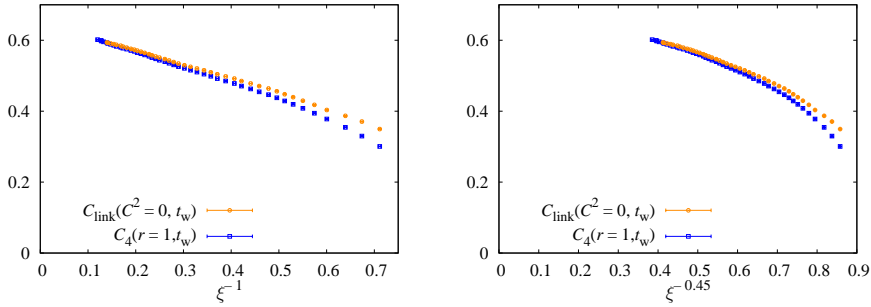}
\caption{Extrapolation of $C_\text{link}(C^2,\tw)$
to $C^2=0$, together with $C_4(r=0,\tw)$, against
$\xi_{1,2}^{-1}(\tw)$ \textbf{(Left)} and $\xi_{1,2}^{-0.45}(\tw)$ \textbf{(Right)}
(plot for our 96 samples at $T=0.6$).}
\label{XI-CLINK-QLINK}
\end{figure}

As for the Edwards-Anderson model, $C_\text{link}$
is seen to be a very smooth function of~$C^2$~\cite{DYNJANUS},
so we can easily compute the curve $C_\text{link}(C^2=0,\tw)$
with a linear extrapolation. We can also study $C_4(r=1,\tw)$,
which is the nonequilibrium disorder average of $Q_\text{link}(\tw)$,
Eq.~\eqref{QLINK}. The two curves are plotted against
$\xi^{-1}(\tw)$ and $\xi^{-0.45}(\tw)$ in Fig.~\ref{XI-CLINK-QLINK}. In accordance with
the previous discussion, and in particular with the
relation $C_\text{link}\bigl(C^2=0,\tw(L)\bigr) = \langle Q_\text{link} | q^2=0\rangle_L$,
both have the same extrapolation to infinite time (they
actually collapse on the same curve for large times). Of the two, the $\xi^{-1}$
scaling is more convincing.

\begin{figure}
\centering
\includegraphics[height=\linewidth,angle=270]{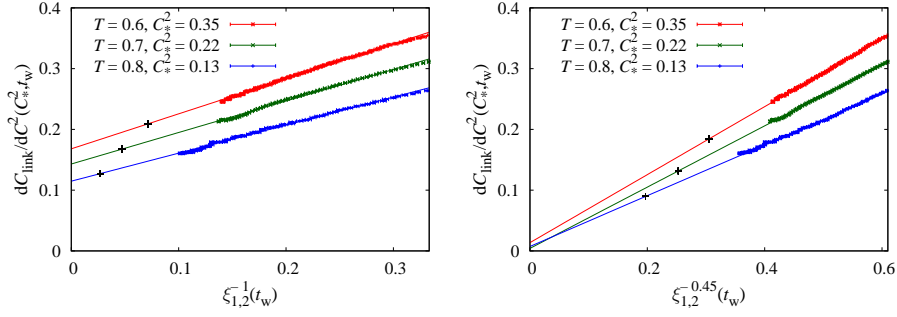}
\caption{Derivative of $C_\text{link}(C^2,\tw)$
with respect to $C^2$  versus
$\xi^{-1}(\tw)$ \textbf{(Left)} and versus
$\xi^{-0.45}(\tw)$ \textbf{(Right)} for
three subcritical temperatures ($T=0.6,0.7,0.8$, from
top to bottom). Lines are
linear least squares fits. We mark by crosses
our extrapolations for
the experimental scale of $\xi(\tw=100\text{ s})$.
The curves are plotted for a fixed value $C^2=C^2_*$, chosen to be just below our lower bound
for $q_\text{EA}$ at each temperature from Eq.~\eqref{COTAS-QEA}.}
\label{PLATEAU}
\end{figure}

Equation~\eqref{CLINKC2-COARSENING} suggests plotting
$\text{d}C_\text{link}(C^2,\tw)/\text{d}C^2$ (see~\cite{DYNJANUS} for
details) against $\xi^{-1}(\tw)$ (Fig.~\ref{PLATEAU}, left) and
against $\xi^{-0.45}(\tw)$ (Fig.~\ref{PLATEAU}, right). It
is important to choose a value $C^2_*$ of $C^2$ smaller than $q_\text{EA}^2$
but not too small, because otherwise the numerical estimate
for the derivative would be unreliable. We have used the most
pessimistically small estimates of $q_\text{EA}^2$ in Table~\ref{TAB-QEA}.
The two representations are linear within our errors. However,
while a $\xi^{-1}(\tw)$ scaling compatible with standard
coarsening seems falsified (the extrapolation to infinite
time is well above zero), a $\xi^{-0.45}(\tw)$ scaling towards zero
is compatible with our data.

\begin{figure}
\centering
\includegraphics[height=\linewidth,angle=270]{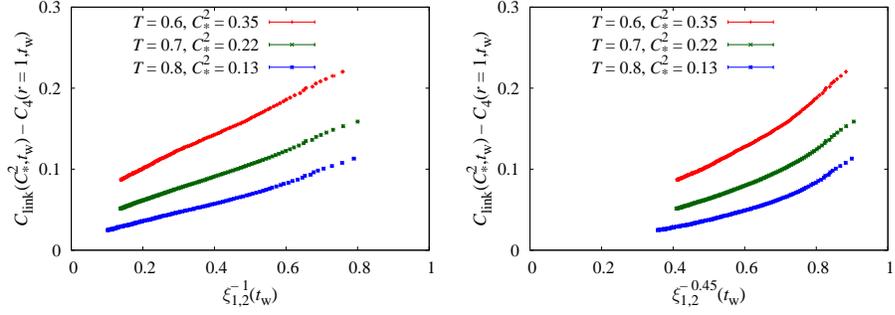}
\caption{Difference $C_\text{link}(C^2_*,\tw)-C_4(r=1,\tw)$
for the same $C^2_*$ of Fig.~\ref{PLATEAU}.
The curves are plotted against $\xi^{-1}(\tw)$ \textbf{(Left)}
and against $\xi^{-0.45}(\tw)$ \textbf{(Right)}. An extrapolation to zero
seems unlikely even for the $\xi^{-0.45}$ case.}
\label{CLINK-QLINK}
\end{figure}

From Fig.~\ref{XI-CLINK-QLINK} and Eq.~\eqref{CLINKC2-COARSENING},
we conclude that the difference $C_\text{link}(C^2,\tw)-C_4(r=1,\tw)$
should vanish as $\xi^{D_\text{s}-D}(\tw)$ in a coarsening system.
We show this difference in Fig.~\ref{CLINK-QLINK},
for the same fixed value $C^2_*$ we used
for the derivative. As we see, the extrapolation as $\xi^{-1}$
is smooth and positive. On the other hand, the $\xi^{-0.45}$
scaling could only be possible if our whole simulation were
in a pre-asymptotic regime.

This discussion notwithstanding, two comments are in order.
First, equation~\eqref{CLINKC2-COARSENING} relies
on equation~\eqref{C-COARSENING}, which is disproved
by the values of exponent $d$ in Table~\ref{TAB-THERMOREMANENT}
(we obtain $d\approx 2$, rather than $d = 3/2$).
Second, we mark by crosses in Fig.~\ref{PLATEAU} the experimentally
relevant scale:  the derivative is certainly nonvanishing
there in either case.

\section{Scaling of the dynamical coherence length}\label{XIGROWTH}
As we have shown in Table~\ref{TAB-a}, our data for the
coherence length is very well represented by a power
law in $\tw$. We found an exponent $z(T)$ roughly
linear in $T^{-1}$. Some theoretical grounds for this behavior
can be found in~\cite{THERMOREMANENT3,zT}. It also appears in the numerical
simulation of the Sherrington-Kirkpatrick model~\cite{zT-NUMERICA}.
Nevertheless, an alternative interpretation has been suggested~\cite{BOUCHAUD-SACLAY}.
We now reanalyze our data under this light.

\subsection{Mixed scaling}
The Saclay group proposed~\cite{BOUCHAUD-SACLAY}, see also~\cite{BERBOU},
a mixed scaling for the dynamical coherence length,
which assumes both critical behavior and activated dynamics in a wide
range of temperatures in the glassy region:
\begin{equation}
\tw \sim \tau_0 \xi^{z_\text{c}} \exp\left(\frac{ Y(T) \xi^\psi}{T}\right)\,,
\label{MS}
\end{equation}
where $\tau_0$ is the microscopical time associated to the dynamics;
$z_\text{c}$ is the dynamical critical exponent computed at the critical
point; $\psi$ is the exponent that takes  the free energy
barriers into account (from the dynamical point of view) and $Y(T)=Y_0
(1-T/T_\mathrm{c})^{\psi \nu}$, with the $\nu$ exponent being the static
critical exponent linked to the coherence length. Near the
critical point $Y(T) \to 0$ and the power law critical dynamics is
recovered.

To asses the validity of the mixed scaling hypothesis, we
consider the following function~\cite{BOUCHAUD-SACLAY}:
\begin{equation}
G(\tw,T)=\left( \frac{ \log(\tw/\tau_0) -z_\text{c} \log
    \xi(\tw,T)}{\xi^\psi T_\text{c}/T}  \right) ^{\frac{1}{\psi \nu}} \,.
\label{GtT}
\end{equation}
Equation (\ref{MS}) would imply that $G(\tw,T)$ is
a $\tw$-independent function of temperature,
\begin{equation}
G(\tw,T)= G_0 \left[1-\frac{T}{T_\mathrm{c}}\right] \,,
\label{G0tT}
\end{equation}
where $G_0=(Y_0/T_\mathrm{c})^{1/(\psi \nu)}$. Both the Ising
and Heisenberg experimental samples%
\footnote{Ag:Mn at $2.5\%$ (Heisenberg like),
                       CdCr$_{1.7}$IN$_{0.3}$S$_4$ (also Heisenberg
                       like) and Fe$_{0.5}$Mn$_{0.5}$TiO$_3$ (Ising like).}
behave consistently with
this expectation. 
In this study, the parameters were taken to be $z_\text{c}=5$
and $\psi=1.5$ (those of Ag\underline{Mn}, a Heisenberg SG)~\cite{BOUCHAUD-SACLAY}.

Notice that $G(\tw,T)$ would be exactly zero if $\xi$ followed a pure power
law with $z$ and $\tau_0$ fixed to their critical temperature values,
\begin{equation}\label{XI-GROWTH-2}
\xi(\tw) = \left[\frac{\tw}{\tau_0(T_\text{c})}\right]^{1/z(T_\text{c})}\ ,
\end{equation}
In the case of a power law with parameters $\tau$ and $z$ different
from those computed at the critical point, we expect that the function
$G(\tw,T)$ should be zero only for $\tw\to\infty$.
In order to avoid a painful and somewhat arbitrary fit, we
take the relevant parameters in Eq.~\eqref{GtT} from the
literature ($z$ and $\tau_0$ are taken at $T_\text{c}$):
$T_\mathrm{c}=1.109$ \cite{PELISSETTO}, $\psi\simeq 0.7$ \cite{KISKER},
$z_\text{c}=6.86$ \cite{DYNJANUS} and  $\nu=2.45$ \cite{PELISSETTO}.

\begin{figure}
\centering
\includegraphics[height=\linewidth,angle=270]{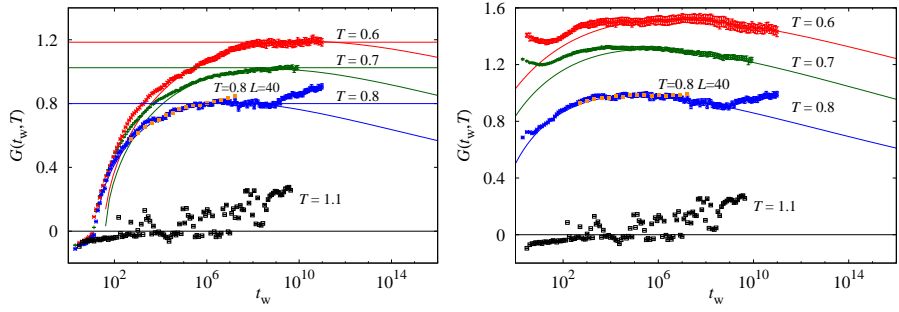}
\caption{\textbf{Left:} Function $G(\tw,T)$ defined in Eq.~\eqref{GtT} versus
$\tw$, for $T=T_\mathrm{c}=1.1$, $T=0.8$, $T=0.7$ and $T=0.6$.
Our estimates for the plateau (see text) are indicated with
horizontal lines. The continuous curves are $G(\tw,T)$
as computed from~\eqref{XI-GROWTH-2}, fixing
$\tau_0(T)=\tau_0(T_\text{c})$, using $z(T)$ computed in the power
law fits of Table~\ref{TAB-a}.
At $T=0.8$ we also show our data for $L=40$.
\textbf{Right:} As in left panel, but now we allow $\tau_0$ to
depend on $T$, Eq.~\eqref{XI-GROWTH-3}.}
\label{FIG_GtT}
\end{figure}

\begin{figure}
\centering
\includegraphics[height=.65\linewidth,angle=270]{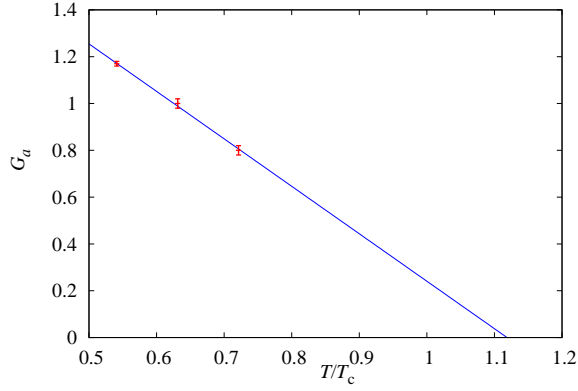}
\caption{The plateaus $G_a(T)$ in Fig.~\ref{FIG_GtT} against
$T/T_\text{c}$. The line is a linear fit ($\chi^2/\text{d.o.f.}=0.12/1$),
that extrapolates to zero at $T/T_\text{c}=1.12(21)$, which is  compatible
with one.}\label{FIG_G0tT}
\end{figure}

In Fig. \ref{FIG_GtT} left, we show the function $G(\tw,T)$ for four
values of the temperature (including $T_\text{c}$).
Although $G(\tw,T)$ is not $\tw$-independent, it plateaus
at a value $G_a(T)$ for long times. This is especially clear for $T=0.6$ and $T=0.8$,
while at $T_\text{c}$ we expect it to be compatible with zero.
For $T=0.7$ the plateau is not well defined, but we estimate it
as the average of the last points \footnote{In order to assign error
  bars to the plateau values, we consider only the biggest contributions,
those from the uncertainties in $z_c$ and in $\tau_0(T_c)$.}. As we show in Fig.~\ref{FIG_G0tT},
$G_a(T)$ behaves consistently with Eq.~\eqref{G0tT}. Hence, we are
in a time region where experimental results for $G(\tw,T)$
are reproduced.

\begin{figure}
\centering
\includegraphics[height=.65\linewidth,angle=270]{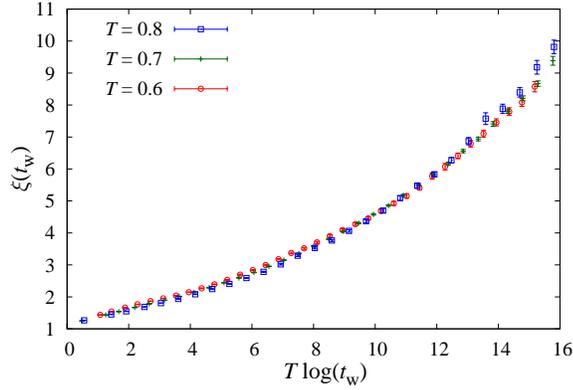}
\caption{Coherence length $\xi_{1,2}(\tw)$ as a function
of $T\log \tw$, for three subcritical temperatures.
Even if the three curves are not equal within errors,
the overall scaling is suggestive.}
\label{XI-LOG}
\end{figure}

However, we will remark the following points:

\begin{enumerate}

\item The departure of the curve for $T=0.8$ from the plateau in Fig.~\ref{FIG_GtT} 
 is a finite size effect, as shown explicitly
by our $L=40$ data (see also~\cite{DYNJANUS}).
A similar, though milder, effect afflicts the data for $T=0.7$
for $\tw> 2.2\times10^{10}$.  The same effect is very clear at $T_\text{c}$,
for even shorter times.

\item Were $\xi(\tw)$ a power law, the curve for $G(\tw,T)$ would
be decreasing (for larger times). On the other hand, we know that finite size
effects cause it to increase. The plateau may well be an
artificial combination of these two effects.
\end{enumerate}

To clear up this problem, we also show in Fig.~\ref{FIG_GtT}, right,
the value of $G(\tw,T)$ computed using the coherence length
and $\tau_0(T)$ from a fit to 
\begin{equation}\label{XI-GROWTH-3}
\xi(\tw) = \left[\frac{\tw}{\tau_0(T)}\right]^{1/z(T)}\ .
\end{equation}
Note that, at variance with Eq.\eqref{XI-GROWTH-2}, we now allow $\tau_0$ to
depend on $T$. These fits were reported in Table~\ref{TAB-a}.
Specifically, we obtained $\tau_0(T=0.6) = 0.008(3)$,
$\tau_0(T=0.7)=0.030(8)$, $\tau_0(T=0.8) = 0.17(4)$
and $\tau_0(T_\text{c})=0.58(13)$.  \footnote{We have found a monotonic
(decreasing) behavior for the microscopic time just as in  Ising samples
(Fe$_{0.5}$Mn$_{0.5}$TiO$_3$)~\cite{SACLAY_EXP}. However, Heisenberg
samples have no such clear pattern: CdCr$_{1.7}$IN$_{0.3}$S$_4$ shows
decreasing monotonic behavior but Ag:Mn at $2.5\%$ and Cu:Mn at 6\% \cite{ORBACH}
present an increasing monotonic one.}

 Although the physical meanings
of Eqs.~\eqref{GtT} and~\eqref{XI-GROWTH-3} are quite different,
the two of them account fairly well for our data.

However, Eq.~\eqref{XI-GROWTH-3}
describes well the data   in nearly the whole range
of times (actually for $\xi\gtrsim 3$)
while Eq.~\eqref{GtT} describes the data only in the region
where the right hand side of Eq.~\eqref{GtT}, computed with Eq.~\eqref{XI-GROWTH-2},
is nearly constant.
Moreover the data for $\xi(\tw)$ nearly collapse if we use the variable
$T\log \tw$, see Fig.~\ref{XI-LOG}. This collapse is not compatible with Eq.~\eqref{MS}.
Henceforth, the activated scaling hypothesis, Eq.~\eqref{MS}, implies
that our data are entirely in a pre-asymptotic regime. We note nevertheless
that extrapolating our data with Eq.~\eqref{XI-GROWTH-3} to the relevant
experimetal scale ($\tw=10^{14}$ or 100 seconds)
produces fairly sensible results~\cite{DYNJANUS}.

\section{Conclusions}\label{CONCLUSIONES}
Using the dedicated computer Janus, we have
studied the nonequilibrium dynamics of the Ising
spin glass for times spanning eleven orders of magnitude.
We have looked into quantities not considered
in our previous work~\cite{DYNJANUS} and  extended
the simulations described therein
by considering more temperatures and vastly enlarging
the number of samples for $T=0.7$.
The emerging picture is that of non-coarsening
dynamics.

We have performed an extensive investigation
of heterogeneous dynamics, by considering
the two-time, two-site correlation function $C_{2+2}(\vn r,t,\tw)$.
We have obtained the first reliable determination
of the nonequilibrium correlation length
and the exponent for the algebraic decay of $C_{2+2}$.
When $t$ is much smaller than $\tw$, the correlation
length reaches a $\tw$-independent limit. On the other hand,
for $t$ much larger than $\tw$, the correlation length
scales as the coherence length. Thus, it might be sensible
to exchange the role of both length scales
in the study of structural glasses~\cite{BIROLI-BOUCHAUD},
where a notion of a coherence length is lacking.

Crucial to the above findings has been our integral
determinations of characteristic length scales.
We have also used them to obtain the coherence length
and to study the replicon mode. Indeed, the exponent
$a$ in Eq.~\eqref{C4-DECAY} is definitively nonvanishing,
and hence incompatible with the droplets picture. We have also considered
nonequilibrium overlap equivalence, with the help
of the coherence length.

We have used both the coherence and correlation lengths
to obtain safe bounds for the Edwards-Anderson
order parameter below the critical temperature.

As for the thermoremanent magnetization,
good agreement with experimental determinations
of the temperature-dependent decay exponents have been obtained. A potentially
useful observation for experimental work is that the thermoremanent magnetization
scales with the coherence length, which
is much harder to measure.
We also observed that a non power law 
function could fit the thermoremanent magnetization
equally well.

The energy relaxation is well described by a power law.
The exponents displayed a nearly linear dependence on temperature.
It has been possible to extrapolate to the critical point,
finding results in nice agreement with the latest
determinations~\cite{DYNJANUS,PELISSETTO}.

We have shown that the link overlap correlation
function $C_\text{link}$ offers a wealth of information
on interphase behavior. Our results have been equally
compatible with the droplets and RSB pictures. However,
in the droplets picture the scaling with the coherence length
of the thermoremanent magnetization is incompatible
with our data. Furthermore, irrespective
of what happens for infinite time,  the variation of
$C_\text{link}$ with $C^2$ is nontrivial at experimental
time scales. This means that the physical view conveyed
by the RSB theory is a better representation
of the physics at the scale of a few hours.

We have critically examined the time growth of the
coherence length, comparing critical and activated
dynamics. We have found that both theories describe
its behavior equally well.

Finally, we have taken the occasion to give
full details of our analysis methods, some of
which are quite new.

\begin{acknowledgements}

We are thankful to P. Nordblad for providing details
on the exponents for thermoremanent magnetization decay.

Janus was supported by EU FEDER funds, with reference
No. UNZA05-33-003 (MEC-DGA, Spain). Janus was developed in
collaboration with ETHlab.  We were partially supported by MEC
(Spain), through contracts No.  FIS2006-08533, FIS2007-60977,
FIS2008-01323, TEC2007-64188; from CAM, contract no. CCG07-UCM/ESP-2532
(Spain) and from the
Microsoft Prize 2007.  D. Yllanes acknowledges support from the FPU
program, reference no. AP2007-01149.

\end{acknowledgements}

\end{document}